\documentclass[dvips]{acta}
\usepackage{supertabular,lscape,epsfig}
\usepackage{amssymb}
\usepackage{amsmath}
\SetPages{1}{1}
\SetVol{63}{2013}

\begin{document}

\begin{Titlepage}
\Title{Variability Survey in the Open Cluster Stock~14 and the Surrounding Fields}
\Author{D.~D~r~o~b~e~k$^1$, ~A.~P~i~g~u~l~s~k~i$^1$, ~R.~R.~S~h~o~b~b~r~o~o~k$^2$ and ~A.~N~a~r~w~i~d$^1$}
{$^1$Astronomical Institute, University of Wroc\l{}aw, Kopernika 11, 51-622 Wroc\l{}aw, Poland\\
$^2$Research School of Astronomy and Astrophysics, Australian National University, Canberra, ACT, Australia\\
E-mail: drobek@astro.uni.wroc.pl}
\Received{Month Day, Year}
\end{Titlepage}

\Abstract{We present the results of a photometric variability survey in the young open cluster Stock~14 and the surrounding
fields. In total, we detected 103 variable stars of which 88 are new discoveries. We confirm short-period, low-amplitude light
variations in two eclipsing members of the cluster, HD~101838 and HD~101794. In addition, we find two new $\beta$~Cephei stars of
which one, HD~101993, is also a member. The sample of pulsating cluster members is supplemented by one multimode slowly pulsating
B-type star and several single-mode candidates of this type. The other pulsating stars in our sample are mostly field stars. In
particular, we found fourteen $\delta$~Scuti stars including one $\gamma$~Dor/$\delta$~Sct hybrid pulsator. From our $UBV$
photometry we derived new parameters of Stock 14: the mean reddening $E(B-V) =$ 0.21\,$\pm$\,0.02~mag, the true distance modulus,
11.90\,$\pm$\,0.05~mag, and the age, 20\,$\pm$\,10~Myr. Finally, we use the new photometry to analyze changes of the 6.322-d
orbital period of the bright eclipsing binary and the member of the cluster, V346~Cen. In addition to the known apsidal motion,
we find that another effect, possibly light-time effect in a hierarchical system of a very long orbital period, affects these
changes. The updated value of the period of apsidal motion for this system amounts to 306\,$\pm$\,4 yr. The open cluster Stock~14
was found to be a fairly good candidate for successful ensemble asteroseismology.}{stars: early type -- stars: oscillations --
stars: individual: HD~101838, HD~101794, HD~101993, V346~Cen -- open clusters and associations: individual: Stock~14}

% ===============================================================================================================================
% ===============================================================================================================================
\section{Introduction}
Asteroseismology is one of a~very few ways allowing to probe stellar interiors. The application of this method to the massive
main-sequence pulsating stars, especially $\beta$~Cephei-type stars, can help to determine the range of convective overshooting
from the core and the core rotation rate (e.g. De~Cat \etal 2011, Lenz 2011). A~prerequisite of successful seismic modeling in
these stars is, however, identification of at least some modes. Many mode identification methods applicable for $\beta$~Cephei
stars were developed during the last two decades, but only some provide reliable results. It is now commonly accepted that
unambiguous mode identification requires precise time-series photometry and spectroscopy supported by careful modeling. However,
this is a~price worth paying. The best attempts of seismic modeling of $\beta$~Cephei stars already allowed to indicate that
stellar cores in massive stars rotate much faster than their envelopes (Dupret \etal 2004, Pamyatnykh \etal 2004, Dziembowski and
Pamyatnykh~2008). Still, there is much more to learn. Asteroseismology gives a~chance of testing stellar physics, especially
stellar opacities that are crucial in driving pulsations in $\beta$~Cephei stars (Pamyatnykh 1999, Miglio \etal 2007, Cugier
2012). Recently a~need for a~further revision of stellar opacities has been pointed out (e.g Zdravkov and Pamyatnykh 2008, Cugier
2012). Such a~revision requires comprehensive theoretical work. It is also clear that in order to have a good material for
comparison, we need to observe more $\beta$~Cephei stars with well-identified modes, spread over the entire instability strip.

So far only several bright $\beta$~Cephei-type stars were modeled in a~comprehensive way. In order to make the results of such
modeling conclusive, parameters of stars under study need to be constrained as well as possible. In this context, there
is a~new possibility which should help to achieve these goals, a method called ensemble asteroseismology. The method can be
applied to pulsating stars in relatively bright open clusters (Suar\'{e}z \etal 2007). It takes advantage of the fact that stars
in a~cluster share the same parameters, namely age, distance and chemical composition. This allows to put many constraints on
parameters of member stars. Additional constraints can be defined if there are bright eclipsing binaries in a~cluster.
Once their masses and radii are obtained from a~combination of the light and radial velocity curves, they can be used to pin down
the isochrones very precisely and use this constraint to derive masses and radii of pulsating stars. Therefore, an open cluster
suitable for ensemble asteroseismology should contain not only many pulsating stars, but preferably also eclipsing binaries. There
is no good example of the application of ensemble asteroseismology for $\beta$~Cephei stars in an open cluster yet, but some work
is in progress and some open clusters, like NGC~6910 (Saesen \etal 2010), were already indicated for this purpose.

It is also clear that new clusters suitable for asteroseismology of massive stars, i.e.~rich in $\beta$~Cephei stars and bright
eclipsing binaries, should be searched for. The present paper is aimed at checking if the open cluster Stock~14 is such an object.
The first indication that Stock~14 might be good for ensemble asteroseismology came from the analysis of time-series photometry
obtained during the third phase of the All Sky Automated Survey (ASAS-3, Pojma\'{n}ski 2002). Using these data, Pigulski and
Pojma\'{n}ski (2008) revealed that two bright members of Stock~14, HD~101794 and HD~101838, appeared to be eclipsing binary
systems with $\beta$~Cephei-type pulsating components. However, because the ASAS-3 telescopes are known to have poor spatial
resolution, it is very important to consider the possibility of photometric contamination of the ASAS-3 photometry by nearby
stars. Such situations have happened before: ALS~1135 is an example (Michalska \etal 2013). For this reason, it was necessary to
perform a~follow-up study to verify if the variability of HD\,101794 and HD\,101838 is both pulsating and eclipsing in nature.
Such a~follow-up study has additional goals: (i) finding new pulsating stars in the cluster, (ii) obtaining multicolor photometry
that can be used in mode identification, and (iii) deriving new parameters of the cluster. In order to achieve these goals, we
obtained new photometric time-series data using a~telescope with spatial resolution superior to that of the ASAS-3 telescopes and
the CCD camera with a~large field of view, amounting to almost half of a~square degree. It is worth noting that our observations
cover a~part of the Cru~OB1 association (Humphreys 1976) that itself can be an interesting object to study.

Stock~14 ($\alpha_{\rm 2000}=\mbox{11}^{\rm{h}}\mbox{43.6}^{\rm{m}}$, $\delta_{\rm 2000} = -\mbox{62}^{\circ}\mbox{31}^{\prime}$)
is a~sparsely populated open cluster of Trumpler type II\,1p (Alter \etal 1970) located close to the Galactic plane
($b\approx-\mbox{0.7}^{\circ}$). Photometric observations of the field of Stock~14 began when L\'{o}den (1973) published results
of photoelectric $UBV$ photometry and spectral types of eleven stars located in the cluster and its vicinity. Stock~14 was
subsequently investigated by Moffat and Vogt (1975), Turner (1982), FitzGerald and Miller (1983), Peterson and FitzGerald (1988),
and McSwain and Gies (2005). The cluster reddening was consistently derived to be of the order of 0.25~mag in terms of $E(B-V)$.
There was some spread in the derived true distance moduli; they ranged from about 11.7~mag (Kharchenko \etal 2005) to 12.3~mag
(FitzGerald and Miller 1983). The presence of early B-type stars distinctly implies a~very young age of the cluster. It was
estimated at 6~Myr by Turner (1982) and 10~Myr by Kharchenko \etal (2005). To this day, no comprehensive study of photometric
variability of stars in the field of Stock~14 has been conducted. Some very early results of this study have been already
published in two conference papers (Drobek \etal 2010, Drobek 2012). In this paper, we present full results of the study.

This paper is organized as follows: first, we describe the instrumentation and the observing run (Sect.~2). Then, calibration and
reduction procedures are outlined. After that (Sect.~3) we describe the steps taken to transform the photometric data to the
standard system, and determine the parameters of the cluster. The main body of the paper is Section 4, in which the variability
survey and its results are presented. The discussion of our results is included in Section~5.

% ===============================================================================================================================
% ===============================================================================================================================
\section{Observations, Calibrations and Reductions}
\begin{figure}[!t]
    \includegraphics[width=\hsize]{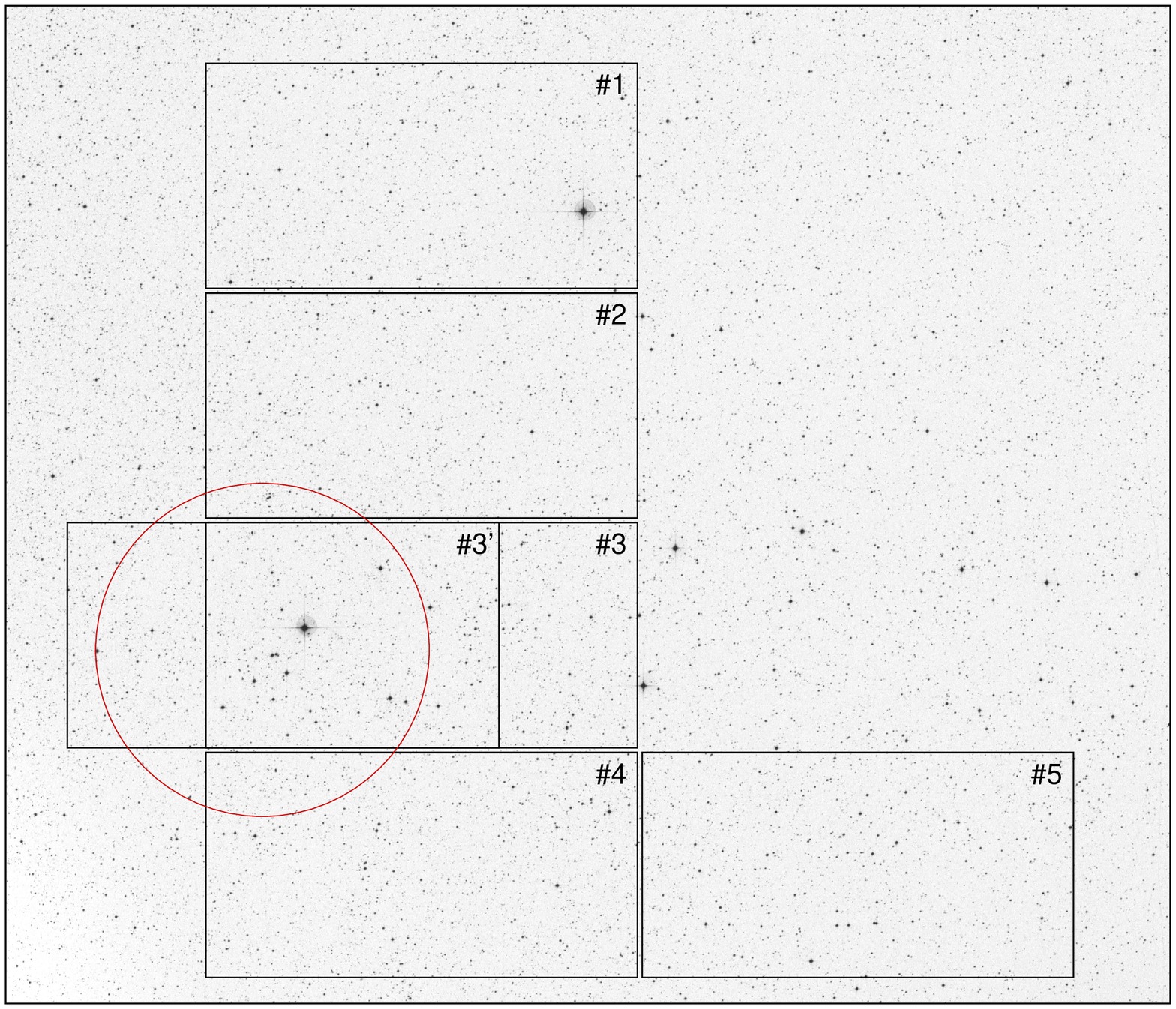}
    \FigCap{The $\mbox{70}^{\prime} \times \mbox{60}^{\prime}$ DSS1 chart centered at $(\alpha_{\rm 2000.0}, \delta_{\rm 2000.0})
=(\mbox{11}^{\rm h}\mbox{41}^{\rm m}, -\mbox{62}^{\circ}\mbox{23}^{\prime})$ showing Stock~14 cluster and the surrounding fields,
with the adopted CCD numbers. North is up, East is to the left. See text for details regarding the area and shape of the field of
view. The circle with a~radius of 10~arcmin is centered at Stock~14.}
    \label{fig:finding_chart}
\end{figure}

The observations of Stock~14 were acquired between March~3 and May~13, 2007, using the Australian National University 1.0-m
telescope at the Siding Spring Observatory, Australia. The detector used for our observations was the Wide Field Imager (WFI),
which was placed in the Cassegrain focus of the telescope. The detector consisted of eight 2\,K\,$\times$\,4\,K CCD chips,
arranged into an 8\,K\,$\times$\,8\,K array. The size of the entire detector was 123\,$\times$\,123~mm, which corresponded to
52\,$\times$\,52~arcmin$^2$ field of view. The resulting image scale of 0.38~arcsec per pixel was vastly superior to that of
ASAS-3 telescopes. Three CCD chips of the WFI were not working at the time of our observing run, so that the field of view was
reduced to about 0.5 square degree. Fig.~\ref{fig:finding_chart} shows the field we observed and numbering system of the CCDs we
adopted.

Our observations consisted of 19 nights in 2007 spanning 71 days and were divided into three separate runs: between March 3 and 11
(first run), between March 22 and April 3 (second run) and between May 8 and 13 (third run). During the first and the third run
only images of the field containing Stock~14 were acquired (CCD \#3$^\prime$, see Fig.~\ref{fig:finding_chart}). In the course of
the second run, the images from all five operational CCDs were attained. Moreover, the entire field of view was shifted
approximately 8~arcmin to the West. A~set of Johnson $UBV$ filters was used. For the field of Stock~14 (CCD \#3) 1168 frames in
$U$, 1169 frames in $B$, and 1193 frames in $V$ filter were obtained. For the remaining fields 655, 657 and 687 frames were
acquired in the respective filters. Since the target stars were bright, typical exposure times were short: 30~s in $U$, 15~s in
$B$ and 10~s in $V$, but varied depending on weather and seeing conditions. On average, seeing amounted to 2.3, 2.2 and 2.0~arcsec
in $U$, $B$ and $V$ filters, respectively. In order to avoid CCD saturation, on nights between March~22 and March~30, the
telescope was deliberately put slightly out of focus. We note that we did not obtain any photometry of the bright stars HD~101570
(CCD \#1) and HD~101947 $=$ V810~Cen (CCD \#3). These stars were severely overexposed in all frames, so that no attempts at
calculating their magnitudes were made.

The calibration of the CCD images consisted of the following steps: overscan subtraction and trimming, bias subtraction,
corrections for non-linearity, and flat fielding. Dark current was found to be negligible. The polynomial non-linearity
corrections were applied according to the prescription of Tinney
(2002)\footnote{\textit{http://www.aao.gov.au/OLDFILES/wfi/wfi\_pfu.html}}. The flat field structure of the WFI CCDs was found to
be variable up to about 2\% in $B$~and $V$,~and up to 5\% in $U$. We decided that the best approach would be to prepare average
master flat frames based on the best available individual frames, and then use the master flats for the images from the entire
observing run. While this approach worked reasonably well for $B$ and $V$ images, we had to discard about 300 worst-looking
$U$~frames.

Stellar magnitudes were calculated by means of the {\sc DAO\-phot~II} program (Stetson 1987). We obtained both the profile-fitting
and aperture photometry. Unfortunately, despite the short exposure times, the brightest stars in our field of view became
saturated, so that it was not possible to obtain their magnitudes by means of aperture photometry. In case of those stars,
the saturated pixels were masked, and profile-fitting photometry was used. For bright unsaturated stars the aperture photometry
appeared to have smaller scatter than profile photometry, while for faint stars the opposite was true. In total, we found 24\,579
stars in our $V$-filter reference frame. To proceed with the analysis, we had to account for the at\-mos\-phe\-ric effects in our
observations. This was done by means of differential photometry. We selected a~set of comparison stars by trial-and-error
procedure, paying attention that they are evenly distributed in the CCD frames. As a~result, we obtained differential time-series
data of each star in up to three passbands.

% ===============================================================================================================================
% ===============================================================================================================================
\section{$UBV$ Photometry}
The mean instrumental magnitude of each star was calculated as an average value of its time-series. The weather and seeing
conditions were optimal during the nights between March~31 and April~3, so that only this part of time-series has been taken into
account for the calculation of average magnitudes. If a~star had less than 40 observations in that time interval, it was
considered too faint and discarded from further analysis. The deviant data points were rejected by means of 3-$\sigma$ clipping.
In eclipsing binaries only the points outside eclipses were considered in that computation. For each star that was observed in
more than one filter we calculated instrumental color indices and their uncertainties. From the inspection of the instrumental
color-magnitude and color-color diagrams, we decided to use the value of 0.01~mag as an upper limit for the uncertainty of the
colors. Using this criterion, we conclude that as far as the instrumental photometry is concerned, from a~total of 24\,579 stars,
a~subset of 6560 has reasonably good photometry in $B$~and $V$ bands, and 1941 stars have decent photometry in three filters.

\subsection{Transformation to the Standard System}
Photometric standards were not observed during our ob\-ser\-ving run. Photoelectric magnitudes of Stock~14 stars were published by
Peterson and FitzGerald (1988), so in principle cluster stars could be used as secondary standards. However, this approach would
only allow to transform the instrumental magnitudes of stars observed with CCD~\#3 (see Fig.~\ref{fig:finding_chart}). The field
of view of the four remaining chips was searched for suitable stars with known $UBV$ magnitudes. Unfortunately, the number of
available stars was too small to derive transformation equations.

During the observations, the field of view of the telescope was drifting to the South-East because of tracking problems. As
a~result, some stars moved from one CCD chip to another during the night. Such stars had their time-series data divided into two
parts, which consisted of data points acquired with two different CCDs. If the number of data points was large enough, it was
possible to calculate average magnitudes of a~star in the instrumental systems defined by both chips. With a~sufficient number
of such stars, photometric transformations between the instrumental systems of adjacent CCDs were derived. Equations of the
following form were used:
\[     v_j = v_i + \alpha_{ij} \times (b-v)_i + \beta_{ij} \]
\[ (b-v)_j =       \gamma_{ij} \times (b-v)_i + \delta_{ij} \]
In all equations and figures, the lowercase and uppercase letters respectively denote the instrumental and standard magnitudes,
while indices stand for the CCD numbers. This convention is used throughout the paper. Values of the coefficients are presented in
Table \ref{tab:inter_ccd}, where the numbers in parentheses are the \textit{rms} errors of the transformation coefficients with
leading zeros omitted. It should be noted that most of the color-dependent terms in the $v$-magnitude transformations
($\alpha_{ij}$ coefficients) are consistent with zero. Derived transformations introduce uncertainties of at most 0.022~mag. The
equations above allow us to bring $v$ magnitudes and $(b-v)$ colors of all stars to the photometric system of CCD~\#3, preparing
them for the transformation to the standard system. Due to the lack of a~suitable number of stars with precise measurements of
$u$~magnitudes near the boundaries of CCDs, the $(u-b)$ colors of stars observed with CCDs other than \#3 have to remain in
their respective instrumental systems.

\MakeTable{c r@{.}l r@{.}l r@{.}l r@{.}l}{12.5cm}{\label{tab:inter_ccd}Coefficients of photometric transformation equations
between adjacent CCD chips of the WFI.}
{
    \hline
    $i \rightarrow j$ & \multicolumn{2}{c}{$\alpha_{ij}$} & \multicolumn{2}{c}{$\beta_{ij}$} &
    \multicolumn{2}{c}{$\gamma_{ij}$} & \multicolumn{2}{c}{$\delta_{ij}$} \\
    \hline
    1 $\rightarrow$ 2 & 0&000(8) & $-$0&838(6)   &  1&013(11)  & $-$0&598(8)  \\
    2 $\rightarrow$ 3 & 0&006(5) &    0&9934(19) &  1&015(12)  &    0&677(5)  \\
    4 $\rightarrow$ 3 & 0&023(6) &    0&487(2)   &  1&09(3)    &    1&018(10) \\
    5 $\rightarrow$ 4 & 0&003(7) & $-$0&555(3)   &  0&925(9)   & $-$0&538(4)  \\
    \hline
}

\begin{figure}[!t]
    \includegraphics[width=\hsize]{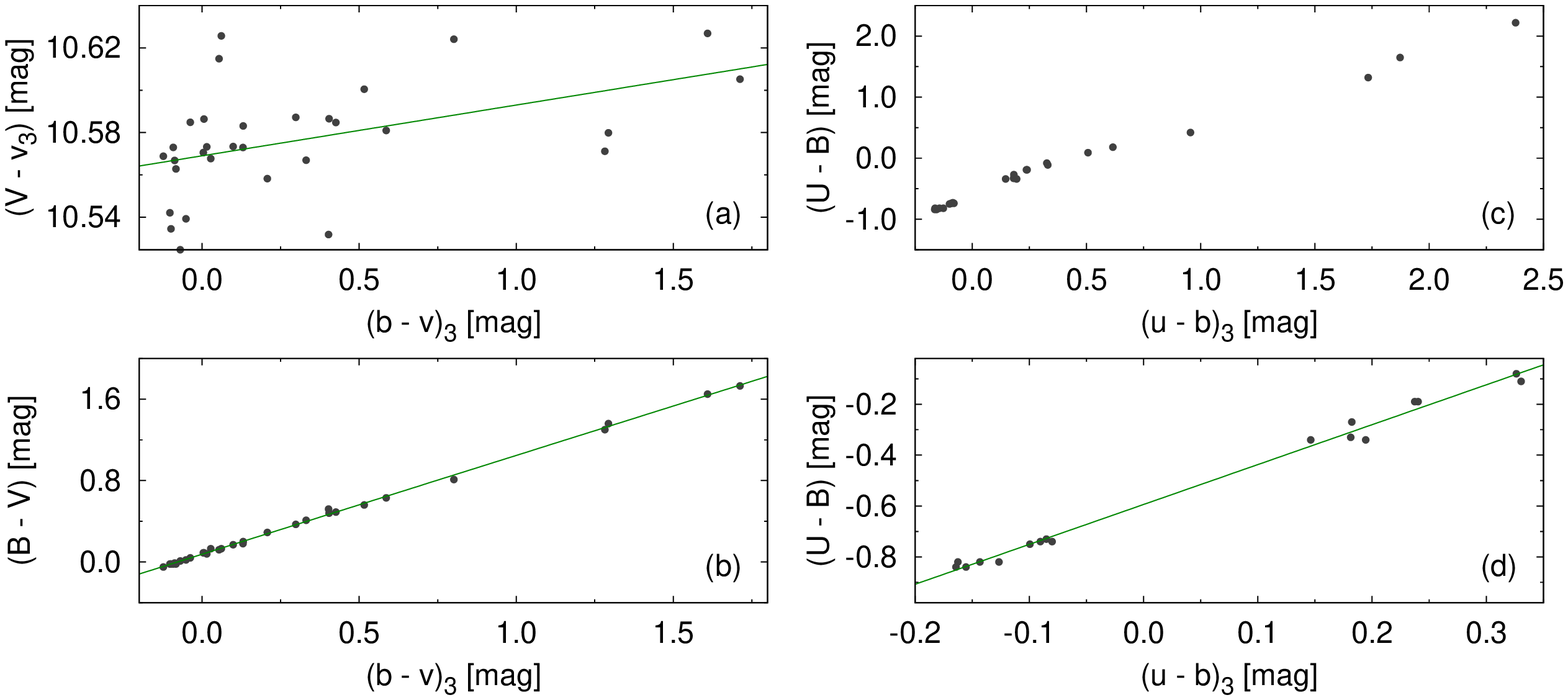}
    \FigCap{Relations between standard and instrumental magnitudes and colors, used to derive photometric transformations. The
transformations of $v_3$ magnitudes and $(b-v)_3$ colors are presented in \textit{panels (a)} and \textit{(b)}, respectively.
\textit{Panel (c)} shows the peculiarities in the $(u-b)_3$ colors, with deviations from linearity for $(u-b)_3 >$ 0.36~mag.
\textit{Panel (d)} shows a~close-up of the range of $(u-b)_3$ colors where the relation to the standard system seems to be
linear.}
    \label{fig:trans_std}
\end{figure}

\begin{figure}[!t]
    \includegraphics[width=\hsize]{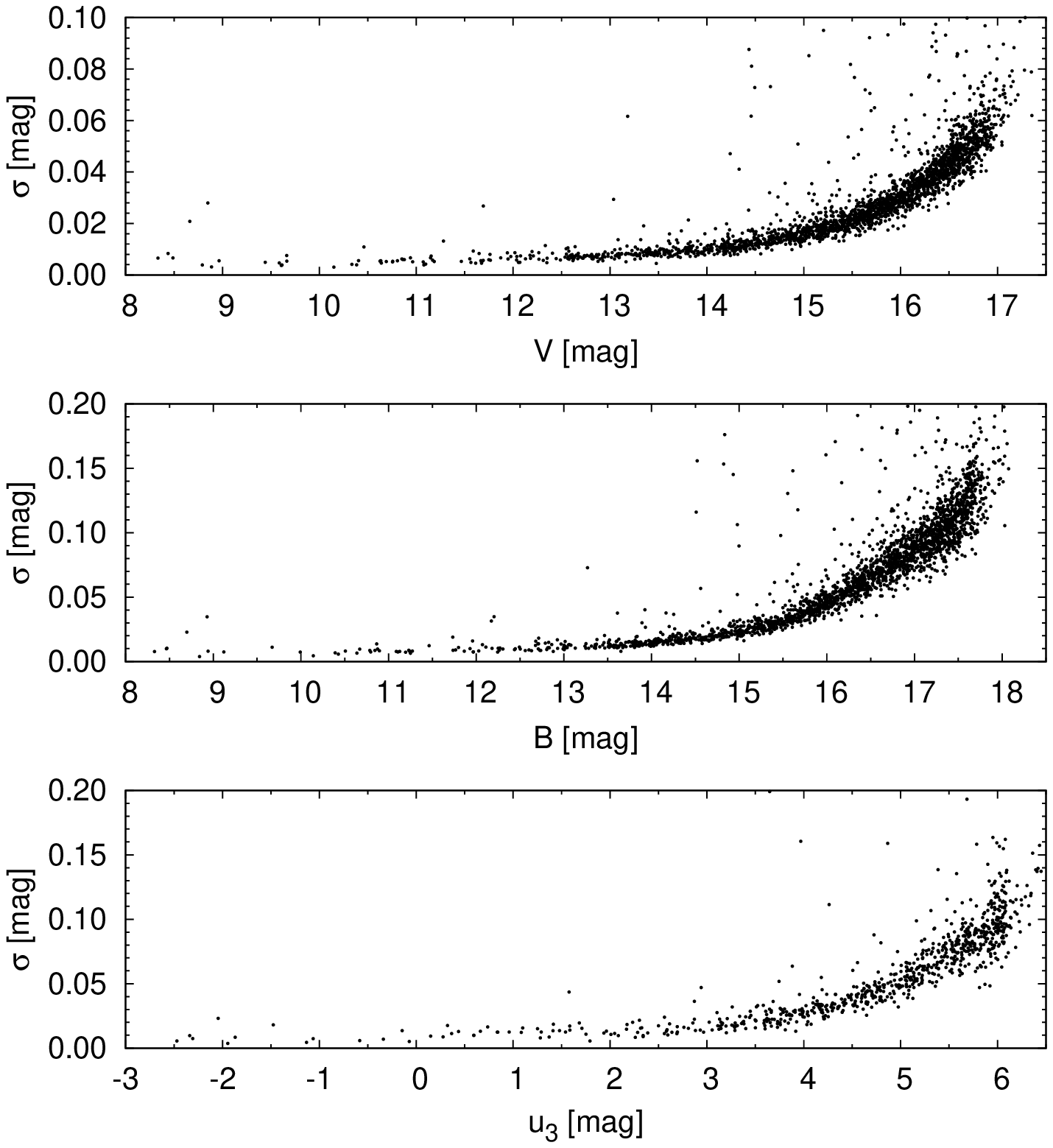}
    \FigCap{The \textit{rms} errors of our $UBV$ photometry plotted as a function of standard ($BV$) and instrumental magnitude
($U$).}
    \label{fig:photo_quality}
\end{figure}

Transformation of the instrumental magnitudes and colors from the instrumental system of CCD~\#3 to the standard system was based
on the photometric data of Peterson and FitzGerald (1988). Transformations of $v_3$ and $(b-v)_3$ are straightforward, and
a~linear function of the instrumental color index fits the 31 data points well (Fig.~\ref{fig:trans_std}a and
Fig.~\ref{fig:trans_std}b). The treatment of $(u-b)_3$ colors required more consideration. From the inspection of
Fig.~\ref{fig:trans_std}c it was clear that a~simple linear transformation would be insufficient for the stars with instrumental
color indices greater than about 0.36~mag. The result of dividing the range of observed color indices into three subsets and
fitting a~line to each of them independently was not satisfactory. The number of stars in the last two subsets was small, which
caused systematic effects in the resulting $(U-B)$ colors. In the end, we decided to derive a~linear transformation based on 17
stars with $(u-b)_3<\mbox{0.36}$~mag (Fig.~\ref{fig:trans_std}d). This transformation was applied to 74 stars from fields 3 and
3$^\prime$. The $(u-b)_3$ colors of the remaining stars were not transformed. The transformation equations in their final forms
are given below, $\sigma$ denotes the standard deviation from the fit:
\[ V = v_3 + \mbox{0.024(8)} \times (b-v)_3 + \mbox{10.569(5)}, \quad \sigma = \mbox{0.024~mag}, \]
\[ (B-V) = \mbox{0.970(6)} \times (b-v)_3 + \mbox{0.076(4)}, \quad \sigma = \mbox{0.018~mag}, \]
\[ (U-B) = \mbox{1.57(3)} \times (u-b)_3 - \mbox{0.594(6)}, \quad \sigma = \mbox{0.026~mag}. \]
The $UBV$ photometry obtained in this way is available from the \textit{Acta Astronomica Archive}. A~plot showing the quality of
the photometry in three filters is presented in Fig.~\ref{fig:photo_quality}. Only the stars observed using CCD~\#3 are shown;
the diagrams are representative of the remaining CCDs.

For illustrative purposes only, a~collective two-color diagram of all the observed stars was prepared. For each CCD chip the
$(u-b)_i$ colors were offset by an average of $(u-b)$ colors of the stars with $(B-V)$ in the range between 0.0 and 0.2~mag. We
denote those offset $(u-b)$ colors as $(u-b)^\prime$. When derived in this manner, the $(u-b)^\prime$ colors of A-type stars have
values close to 0.0~mag. While this procedure introduces some additional scattering of data points to the diagram, it allows to
directly compare the stars observed with different CCD chips of the WFI.

It is clear that the photometric performance of the WFI is unusual and requires an explanation. Tinney
(2002)\footnote{\textit{http://www.aao.gov.au/OLDFILES/wfi/wfi\_pfu.html}} presents the quantum efficiency curves of individual
CCD chips, as well as the transmission curves of the filters. The instrumental $U$ filter seems to be similar to the standard one,
but the major issue is the quantum efficiency. The WFI CCDs are optimized for observation in the visual and red parts of the
spectrum, and their quantum efficiency drops to less than 10\% on the ultraviolet side of the Balmer jump. This shifts the
effective wavelength of the $U$ filter to longer wavelengths, possibly to the red side of the Balmer jump. Additionally, the
transmission curve of the WFI $B$ filter is shifted towards shorter wavelengths in comparison to the standard Johnson $B$. We
believe the combination of these two factors to be the cause of the peculiar behavior of the instrumental $(u-b)$ colors.

\subsection{The Parameters of the Cluster}
\begin{figure}[!t]
    \includegraphics[width=\hsize]{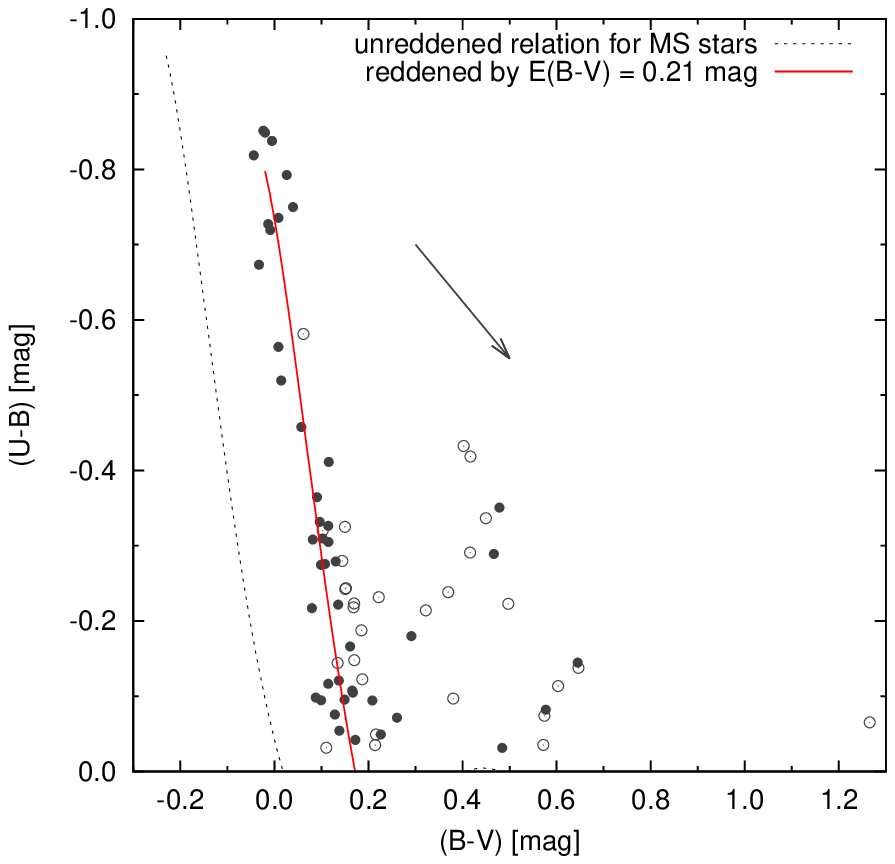}
    \FigCap{Determination of the reddening of Stock~14. Dotted line represents the unreddened relation of Caldwell \etal (1993),
solid line represents the same relation reddened by $E(B-V)=\mbox{0.21}$~mag. The arrow is a~reddening vector corresponding to the
derived color excess. Filled symbols correspond to stars located within 10 arcmin from the center of the cluster, and open symbols
denote the remaining stars. Only 74 stars with reliable $(U-B)$ color indices are shown.}
    \label{fig:fit_caldwell}
\end{figure}

Having calculated standard magnitudes and color indices of cluster stars, it was possible to determine the global parameters of
Stock~14. To measure the reddening of the cluster, a~standard relation between intrinsic color indices of main sequence stars
(Caldwell \etal 1993) was fitted to the two-color diagram of the cluster (Fig.~\ref{fig:fit_caldwell}). Only 74 stars with
reliable $(U-B)$ color indices have been taken into account. The stars have been divided into two sets: stars located within 10
arcmin from the center of the cluster (plotted with filled symbols), and the remaining stars (open symbols). We assume that the
first set represents more probable cluster members, although from the inspection of Fig.~\ref{fig:fit_caldwell} it is clear that
there are some stars with individual reddenings concordant with the one of the cluster located further than 10 arcmin from the
center of Stock~14. We assumed a~standard relation between color excesses, $E(U-B)/E(B-V)=\mbox{0.72} + \mbox{0.05} \times
E(B-V)$, and fitted the reddened curve to 55 probable cluster members. The reddening of $E(B-V)=\mbox{0.21} \pm \mbox{0.02}$~mag
was determined.

\begin{figure}[!t]
    \includegraphics[width=\hsize]{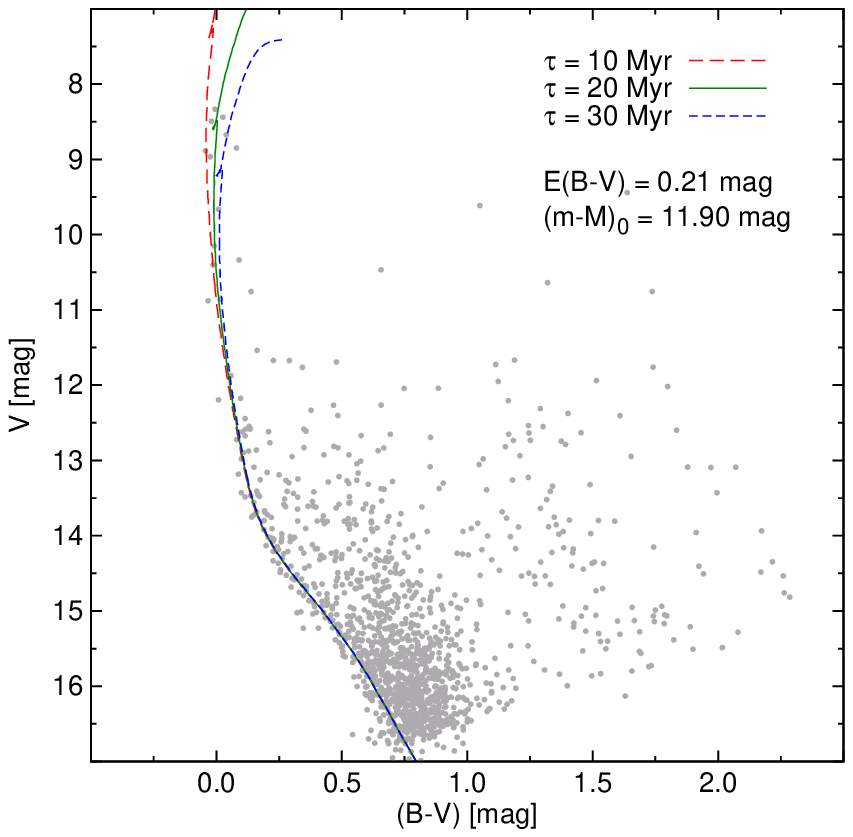}
    \FigCap{The color-magnitude diagram of Stock~14 with the isochrones from Bertelli \etal (2008, 2009) fitted to the main
sequence of the cluster. The best-fitting isochrone corresponds to the true distance modulus $(m-M)_0=\mbox{11.90} \pm
\mbox{0.05}$~mag and the age $\tau=\mbox{20} \pm \mbox{10}$~Myr. Solar metallicity was assumed.}
    \label{fig:cmd_isofit}
\end{figure}

Once the reddening of the cluster has been estimated, distance and age of Stock~14 were determined by fitting theoretical
isochrones to the color-magnitude diagram of cluster stars (Fig.~\ref{fig:cmd_isofit}). In this case the number of available
stars with reliable standard photometry was much larger. To clarify the picture, only the stars located within the radius of
10~arcmin from the cluster have been used. Despite this limitation, the diagram is dominated by the main sequence field stars.
Nevertheless, a~thin, well-defined main sequence of Stock~14 is visible down to $V$ magnitude of about 15.5. We note the apparent
lack of evolved stars in this region of the sky, save for V810~Cen for which we have no photometry due to saturation, and which
probably is not phy\-si\-cally associated with Stock~14 anyway (Kienzle \etal 1998). To derive the age and distance of the cluster
we used a~set of theoretical solar-metallicity ($Z=\mbox{0.017}$) isochrones of Bertelli \etal (2008, 2009). They were fitted to
the main sequence of Stock~14. Assuming $A_V/E(B-V)=\mbox{3.1}$, we derived the true distance modulus of $(m-M)_0=\mbox{11.90} \pm
\mbox{0.05}$~mag. The age of the cluster amounts to 20 $\pm$ 10~Myr (Fig.~\ref{fig:cmd_isofit}). Since previous estimates of age
and distance of Stock~14 done by different authors were based on smaller samples of stars, the results presented here can be
regarded as more reliable.

% ===============================================================================================================================
% ===============================================================================================================================
\section{Variable Stars}
We decided that with up to 1200 observations per star, the $V$-filter profile photometry is best suited for the variability
survey. The differential $V$-band time-series were subjected to the Fourier analysis in the frequency range between 0~{d}$^{-1}$
and 40~{d}$^{-1}$. The resulting Fourier spectra and phase diagrams were examined visually. At that point the decision on
classification was made. If a~star was classified as variable, photometry in the remaining passbands was inspected. If a~star was
bright and unsaturated, aperture photometry data were also examined. In the final ana\-ly\-sis, we used the best available
time-series data for each star. Whenever a~frequency was found in the power spectrum, the following function was fitted to the
original data by means of the non-linear least squares:
\[m(t) = m_0 + \sum_{i=1}^{N} A_i \sin(2\pi f_i t +\phi_i)\]
where $m_0$ is the mean differential magnitude, $N$ is the number of terms, $f_i$, $A_i$ and $\phi_i$ are their frequencies,
amplitudes and phases, and $t$ is the time elapsed from the initial epoch of $T_0 = \mbox{HJD\,2\,453\,000.0}$. Then, the power
spectrum of the residuals was calculated. This procedure was repeated until there were no more significant peaks in the spectrum.
In terms of the signal-to-noise ratio, we adopted a~significance threshold of 4.0. When a~star appeared to be an eclipsing binary
system, we used the AoV method of Schwarzenberg-Czerny (1996) as a~consistency check, utilising an appropriate number of harmonics
to derive the orbital period. If a~significant frequency below 4~d$^{-1}$ was detected in a~given star, its differential
time-series was recalculated using a~new set of suitable comparison stars located near that star in the CCD frame. This was done
to ascertain that the low frequency does not result from the inaccuracies of flat-fielding. The amplitude spectra of some stars
exhibited noise at low frequencies. For those stars we used the following detrending procedure: the periodic signal was removed
from the original time-series, and the average of residuals on each night was calculated. Those averages were subsequently
interpolated by a~cubic spline, and that smooth curve was subtracted from the original data.

\begin{figure}[!t]
    \includegraphics[width=\hsize]{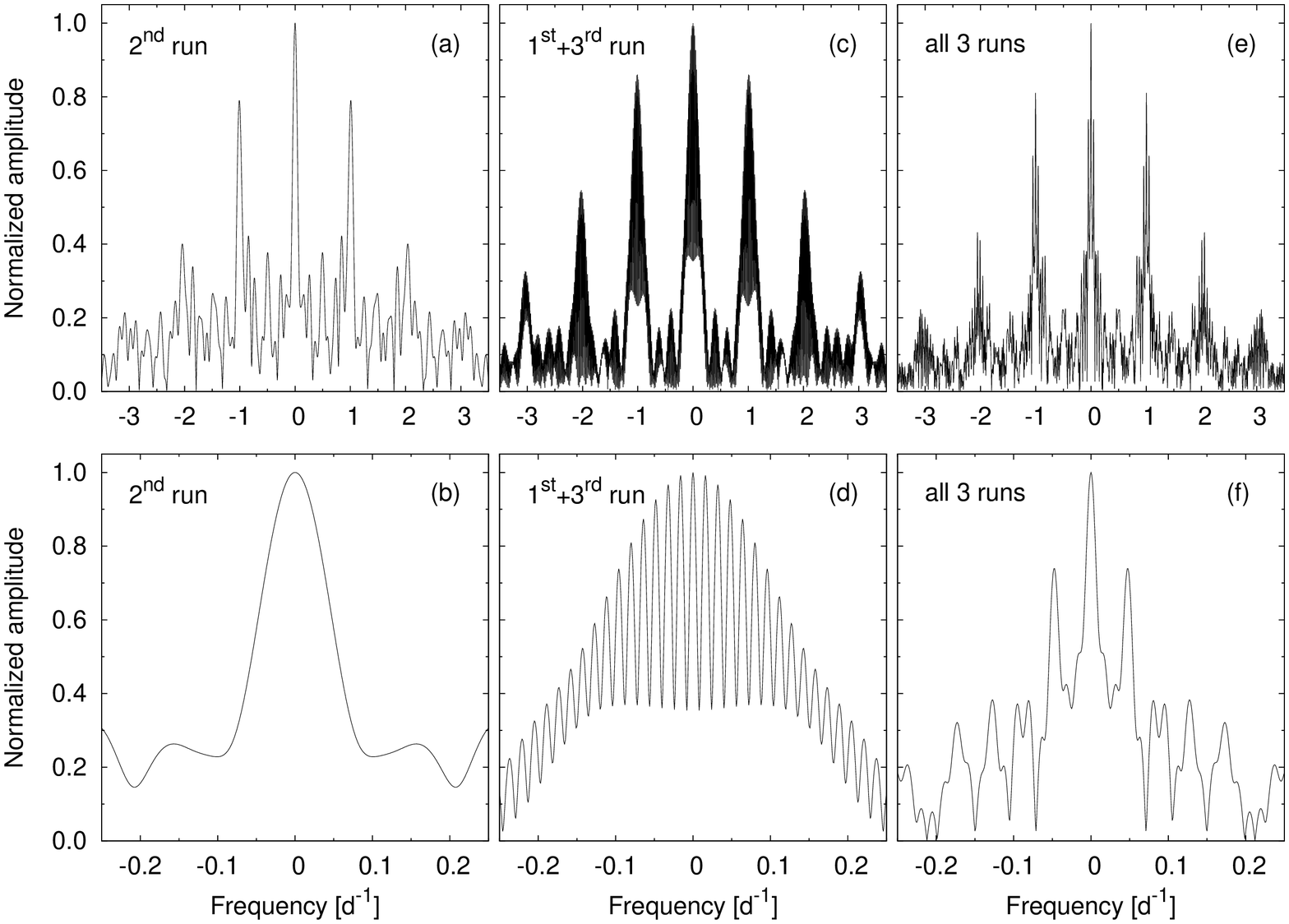}
    \FigCap{Spectral windows of our observations. \textit{Panels (a)} and \textit{(b)}: for objects observed during the second
observing run only; \textit{(c)} and \textit{(d)}: for objects observed during the first and third run; \textit{(e)} and
\textit{(f)}: for objects observed during all three runs. \textit{Panels (b)}, \textit{(d)} and \textit{(f)} show close-ups on the
central peaks in the frequency spectra. See text for details.}
    \label{fig:spectral_window}
\end{figure}

Because our data set consists of observations acquired during three observing runs, and because the field of view was not the same
during all the runs, we encounter three types of spectral windows (Fig.~\ref{fig:spectral_window}). Stars observed in fields 1, 2,
4, 5 and the westernmost part of field 3 (see Fig.~\ref{fig:finding_chart}) were observed during the second run only, and the
spectral windows of their observations are presented in Fig.~\ref{fig:spectral_window}a and Fig.~\ref{fig:spectral_window}b. Stars
located in the easternmost part of field 3$^\prime$ were observed during the first and third run (Fig.~\ref{fig:spectral_window}c
and Fig.~\ref{fig:spectral_window}d). Finally, the region of the sky closest to the center of Stock~14 was observed during all
runs (Fig.~\ref{fig:spectral_window}e and Fig.~\ref{fig:spectral_window}f). The frequency resolution is the best for the combined
data, but an aliasing problem is introduced: because the three runs are separated by about 20~days (as described in Section~2),
the spectral window features strong aliases of frequency $\pm$\,0.048~d$^{-1}$ (Fig.~\ref{fig:spectral_window}f).

For sufficiently bright and isolated stars with high enough light amplitudes, we used all the available ASAS-3 photometric data to
derive precise frequencies. While the accuracy of the ASAS-3 photometry is inferior to the one obtained by us, its frequency
resolution is far superior to ours. This is because the time base of ASAS-3 observations spans up to 10 years, depending on
a~star. Admittedly, the ASAS-3 frequency spectra are affected with severe daily aliases, but it is possible to calculate the first
approximation of the frequency from the WFI photometry, and use this value to discard spurious results in the ASAS-3 periodogram.
Whenever this paper mentions a~period or frequency derived from the ASAS-3 data without any citations, it indicates that it has
been derived in this manner.

\MakeOwnTable{rccr@{.}l r@{.}l r@{.}l  r@{.}l r@{.}ll}{\hsize}{\label{tab:var_stars}
Variable stars in the field of Stock 14 open cluster. Variable stars which were known prior to our study are preceded with an
asterisk in the first column. Likely members of Stock 14 are identified by an (m) following the identification. $\Delta V$ stands
for the peak-to-peak amplitude. Periods followed by [A] have been calculated from the ASAS-3 photometry. The other columns are
self-explanatory.}{\tiny}
{\hline\noalign{\smallskip}
 ID  & $\alpha_{2000}$ & $\delta_{2000}$ & \multicolumn{2}{c}{$V$} & \multicolumn{2}{c}{$B-V$} &
\multicolumn{2}{c}{$(u-b)^\prime$}& \multicolumn{2}{c}{$\Delta V$} & \multicolumn{2}{c}{$P$} &
Variability type, remarks\\
     & [$^{\rm{h~m~s}}$] & [$^{\rm{o}~\prime~\prime\prime}$] & \multicolumn{2}{c}{[mag]} & \multicolumn{2}{c}{[mag]} &
\multicolumn{2}{c}{[mag]} & \multicolumn{2}{c}{[mag]} & \multicolumn{2}{c}{[d]} &\\
\noalign{\smallskip}\hline
$\ast$ v1 (m)& 11 42 49.32 & $-$62 33 55.0 & 8&439 & 0&026 & $-$0&713 & 0&024 & 0&31976(4) & $\beta$~Cep (HD\,101838, W\,14)\\
   &&&\multicolumn{2}{c}{}&\multicolumn{2}{c}{}&\multicolumn{2}{c}{}& 0&029 & 5&41166(7) [A] & + EA\\
$\ast$ v2 (m)& 11 42 25.37 & $-$62 28 37.5 & 8&673 & 0&039 & $-$0&686 & 0&037 & 0&224685(19) & $\beta$~Cep (HD\,101794, W\,13)\\
   &&&\multicolumn{2}{c}{}&\multicolumn{2}{c}{}&\multicolumn{2}{c}{}& 0&079 & 1&4632361(21) [A] & + EA\\
   v3 (m) & 11 43 46.83 & $-$62 30 49.4 &  9&663 & 0&008 & $-$0&676 & 0&011 & 0&084724(4) & $\beta$~Cep (HD\,101993, W\,2)\\
   v4     & 11 40 59.58 & $-$62 00 14.9 & 13&639 & 0&866 & $-$0&255 & 0&027 & 0&24104(26) & $\beta$~Cep\\
\noalign{\smallskip}\hline\noalign{\smallskip}
   v5 (m) & 11 41 48.47 & $-$62 04 01.1 & 11&229 & 0&052 & $-$0&558 & 0&048 & 2&133(10)        & SPB candidate\\
$\ast$ v6 & 11 41 41.00 & $-$62 36 26.8 & 11&280 & 0&450 & $-$0&422 & 0&038 & 0&658547(10) [A] & SPB candidate\\
   v7     & 11 42 05.96 & $-$62 31 17.2 & 11&587 & 0&369 & $-$0&359 & 0&038 & 0&328955(22)     & SPB candidate (W\,421)\\
   v8 (m) & 11 40 00.35 & $-$62 44 17.9 & 11&759 & 0&032 & $-$0&579 & 0&086 & 1&8253(23)       & SPB candidate\\
   v9 (m) & 11 39 37.86 & $-$62 48 10.8 & 12&189 & 0&022 & $-$0&476 & 0&014 & 0&6127(16)       & SPB candidate\\
  v10 (m) & 11 44 11.80 & $-$62 31 59.3 & 12&195 & 0&009 & $-$0&567 & 0&074 & 3&4108(13)       & SPB candidate (W\,7)\\
  v11 (m) & 11 43 24.56 & $-$62 34 16.1 & 12&547 & 0&130 & $-$0&385 & 0&033 & 1&1225(4)        & SPB (W\,163 = W\,441)\\
  v12 & 11 39 23.05 & $-$62 38 33.3 & 12&871 & 0&356 &    0&078  & 0&018 & 0&3285(6)  & multiperiodic \\
  v13 & 11 39 33.63 & $-$62 44 34.0 & 13&032 & 0&411 & $-$0&056  & 0&062 & 0&6835(7)  & multiperiodic\\
  v14 & 11 42 58.29 & $-$62 07 20.6 & 14&185 & 0&623 & $-$0&093  & 0&039 & 0&5149(10) & SPB candidate\\
  v15 & 11 44 11.63 & $-$62 40 29.1 & 15&203 & 0&611 & \multicolumn{2}{c}{---} & 0&036 & 0&876(3)   & multiperiodic\\
\noalign{\smallskip}\hline\noalign{\smallskip}
  v16 & 11 41 48.14 & $-$62 39 42.6 & 10&877 & 0&324 & $-$0&002  & 0&042 & 0&22247(9)    & $\delta$~Sct: (HD\,308900)\\
  v17 & 11 42 13.95 & $-$62 15 39.4 & 12&747 & 0&485 & $-$0&009  & 0&030 & 0&11688(4)    & $\delta$~Sct\\
  v18 & 11 41 34.04 & $-$62 17 53.8 & 12&927 & 0&496 &    0&016  & 0&035 & 0&10555(4)    & $\delta$~Sct\\
  v19 & 11 43 01.88 & $-$61 59 09.6 & 13&089 & 0&536 &    0&034  & 0&030 & 0&069587(19)  & $\delta$~Sct\\
  v20 & 11 41 36.89 & $-$62 07 02.9 & 13&407 & 0&371 &    0&006  & 0&024 & 0&053869(14)  & $\delta$~Sct\\
  v21 & 11 40 52.37 & $-$62 38 25.2 & 13&478 & 0&356 &    0&219  & 0&019 & 0&4660(14)    & $\delta$~Sct / $\gamma$~Dor
hybrid\\
  v22 & 11 41 59.68 & $-$62 41 54.3 & 13&502 & 0&461 &    0&103  & 0&035 & 0&092018(20)  & $\delta$~Sct\\
  v23 & 11 37 25.71 & $-$62 41 45.2 & 13&529 & 0&326 & $-$0&065  & 0&023 & 0&071982(23)  & $\delta$~Sct\\
  v24 & 11 45 00.03 & $-$62 31 56.9 & 13&809 & 0&442 & $-$0&004  & 0&047 & 0&127402(4)   & $\delta$~Sct\\
  v25 & 11 41 45.82 & $-$62 48 49.0 & 13&894 & 2&004 & \multicolumn{2}{c}{---} & 0&022 & 0&2482(13)    & multiperiodic\\
  v26 & 11 41 26.12 & $-$62 02 36.2 & 14&376 & 0&492 & $-$0&020  & 0&027 & 0&07072(3)    & $\delta$~Sct\\
  v27 & 11 42 25.28 & $-$62 32 56.4 & 14&931 & 0&763 &    0&219  & 0&046 & 0&0740460(24) & $\delta$~Sct\\
  v28 & 11 41 03.02 & $-$62 48 38.5 & 14&995 & 0&534 & \multicolumn{2}{c}{---} & 0&027 & 0&051441(15)  & $\delta$~Sct\\
  v29 & 11 43 58.86 & $-$62 45 28.1 & 15&027 & 0&527 & \multicolumn{2}{c}{---} & 0&035 & 0&08317(3)    & $\delta$~Sct\\
  v30 & 11 42 04.12 & $-$62 21 03.5 & 16&069 & 1&134 & \multicolumn{2}{c}{---} & 0&045 & 0&08668(4)    & $\delta$~Sct\\
\noalign{\smallskip}\hline\noalign{\smallskip}
$\ast$ v31 & 11 40 58.58 & $-$62 41 33.0 & 8&724 & 0&827 & 0&398 & 0&836 & 3&334311(3) [A] & Cepheid (HD\,101602, \\
\multicolumn{13}{c}{}&UZ~Cen)\\
$\ast$ v32 & 11 41 19.48 & $-$62 15 53.2 &12&953 & 1&954 & 0&881 & 0&450 & 5&89237(21) [A] & Cepheid (IZ~Cen)\\
\noalign{\smallskip}\hline\noalign{\smallskip}
$\ast$ v33 & 11 44 16.61 & $-$62 33 47.4 & 8&849 & 0&080 &$-$0&346& 0&085 &11&3638(4) [A] & $\alpha^2$~CVn (HD\,102053, W\,7)\\
  v34 & 11 37 30.37 & $-$62 48 07.6 & 12&813  &     0&681 &     0&050 & 0&035  &  1&1044(26)     & monoperiodic\\
  v35 & 11 38 26.42 & $-$62 44 58.5 & 13&512  &     0&431 &     0&044 & 0&021  &  0&23089(21)    & monoperiodic\\
  v36 & 11 38 45.88 & $-$62 39 12.7 & 13&630  &     0&676 &     0&129 & 0&055  &  1&550(5)       & monoperiodic\\
  v37 & 11 42 12.67 & $-$62 44 43.7 & 13&861  &     0&845 & \multicolumn{2}{c}{---} & 0&063  &  3&808(24)    & monoperiodic\\
  v38 & 11 40 43.27 & $-$62 38 42.6 & 13&973  &     0&520 & \multicolumn{2}{c}{---} & 0&017  &  0&13306(11)  & monoperiodic\\
  v39 & 11 44 03.29 & $-$62 18 32.1 & 14&515  &  1&365 & \multicolumn{2}{c}{---} & 0&098  & 1&2579(18) & monoperiodic,
non-sinusoidal\\
  v40 & 11 37 51.02 & $-$62 44 24.5 & 14&597  &     1&118 & \multicolumn{2}{c}{---} & 0&140  &  0&138916(21) & monoperiodic\\
  v41 & 11 42 36.96 & $-$62 22 20.1 & 14&981  &     0&604 &     0&192 & 0&030  &  0&2732(4)      & monoperiodic \\
  v42 & 11 43 26.63 & $-$62 36 09.1 & 15&959  &     0&681 & \multicolumn{2}{c}{---} & 0&096  &  0&189739(13) & monoperiodic\\
  v43 & 11 42 13.99 & $-$62 28 50.4 & 16&600  &     0&768 & \multicolumn{2}{c}{---} & 0&112  &  0&187150(16) & monoperiodic\\
  v44 & 11 42 53.03 & $-$62 20 25.6 & 16&604  & \multicolumn{2}{c}{---} & \multicolumn{2}{c}{---} & 0&188  &  0&7936(17)   &
monoperiodic\\
  v45 & 11 41 27.38 & $-$62 08 33.3 & 16&728  & \multicolumn{2}{c}{---} & \multicolumn{2}{c}{---} & 0&172  &  0&23455(23)  &
monoperiodic\\
  v46 & 11 41 48.78 & $-$62 19 20.6 & 16&924  & \multicolumn{2}{c}{---} & \multicolumn{2}{c}{---} & 0&234  &  0&14170(25)  &
monoperiodic\\
  v47 & 11 41 30.64 & $-$62 10 32.9 & 16&930  & \multicolumn{2}{c}{---} & \multicolumn{2}{c}{---} & 0&202  &  0&2188(3)    &
monoperiodic\\
  v48 & 11 41 59.05 & $-$62 28 06.5 & 16&969  & \multicolumn{2}{c}{---} & \multicolumn{2}{c}{---} & 0&178  &  0&255421(23) &
monoperiodic\\
  v49 & 11 44 03.71 & $-$62 45 33.3 & 17&110  & \multicolumn{2}{c}{---} & \multicolumn{2}{c}{---} & 0&169  &  0&4686(8)    &
monoperiodic\\
  v50 & 11 44 07.53 & $-$62 22 36.3 & 17&218  & \multicolumn{2}{c}{---} & \multicolumn{2}{c}{---} & 0&217  &  0&13490(9)   &
monoperiodic\\
  v51 & 11 43 38.42 & $-$62 18 50.0 & 17&648  & \multicolumn{2}{c}{---} & \multicolumn{2}{c}{---} & 0&111  &  0&12133(11)  &
monoperiodic\\
  v52 & 11 43 47.13 & $-$62 23 14.2 & 17&770  & \multicolumn{2}{c}{---} & \multicolumn{2}{c}{---} & 0&499  &  0&23121(17)  &
monoperiodic\\
  v53 & 11 43 08.12 & $-$62 23 29.6 & 17&953  & \multicolumn{2}{c}{---} & \multicolumn{2}{c}{---} & 0&188  &  0&12360(14)  &
monoperiodic\\
  v54 & 11 41 22.56 & $-$62 15 55.1 & 18&403  & \multicolumn{2}{c}{---} & \multicolumn{2}{c}{---} & 0&170  &  0&10023(9)   &
monoperiodic\\
    \noalign{\smallskip}\hline
}

\setcounter{table}{1}
\MakeOwnTable{rccr@{.}l r@{.}l r@{.}l  r@{.}l r@{.}ll}{\hsize}{Concluded}{\tiny}
{\hline\noalign{\smallskip}
 ID  & $\alpha_{2000}$ & $\delta_{2000}$ & \multicolumn{2}{c}{$V$} & \multicolumn{2}{c}{$B-V$} &
\multicolumn{2}{c}{$(u-b)^\prime$}& \multicolumn{2}{c}{$\Delta V$} & \multicolumn{2}{c}{$P$} &
Variability type, remarks\\
     & [$^{\rm{h~m~s}}$] & [$^{\rm{o}~\prime~\prime\prime}$] & \multicolumn{2}{c}{[mag]} & \multicolumn{2}{c}{[mag]} &
\multicolumn{2}{c}{[mag]} & \multicolumn{2}{c}{[mag]} & \multicolumn{2}{c}{[d]} &\\
\noalign{\smallskip}\hline\noalign{\smallskip}
$\ast$ v55 (m) & 11 42 49.70 & $-$62 26 05.5 &  8&492  & $-$0&020 & $-$0&749 &   0&291  &  6&321900(12) [A]   &  EA (HD\,101837,
V346~Cen,\\
\multicolumn{13}{c}{}&W\,12)\\
$\ast$ v56 & 11 42 56.41 & $-$62 48 24.4 &10&185  &      0&345 & $-$0&570 &   0&039  &  0&8921687(24) [A] &  EA (HD\,309018)\\
  v57 & 11 44 41.60 & $-$62 33 17.9 & 11&692  &    0&478 & $-$0&431 &  0&095  &  3&36366(8) [A]   & Ellipsoidal (W\,434)\\
  v58 & 11 42 01.89 & $-$62 33 16.3 & 12&037  &    0&214 & $-$0&230 &   0&281  &  1&63343(4)      & EA (W\,433)\\
  v59 & 11 43 18.74 & $-$62 41 27.0 & 12&077  &    0&233 & $-$0&154 &   0&182  &  1&285701(13)    & EA\\
  v60 & 11 42 06.66 & $-$62 16 02.2 & 12&409  &    0&496 & $-$0&022 &   0&164 &  0&684227(4) [A]  & EA + EA\\
  v61 & 11 44 35.66 & $-$62 33 50.2 & 13&010  &    0&575 & 0&023 &  0&082  &  0&437418(12)        & EW (W\,438)\\
  v62 & 11 40 45.44 & $-$62 07 39.4 & 13&135  &    0&495 & 0&066 &  0&218  &  0&64738(21)         & EB\\
  v63 & 11 37 19.29 & $-$62 45 53.9 & 13&985  &    0&433 & $-$0&055 &  0&048  &  0&5950(5)        & Ellipsoidal\\
  v64 & 11 40 48.40 & $-$62 18 46.2 & 14&257  &    0&397 & 0&094 &  0&060  &  0&5835(5)           & Ellipsoidal\\
  v65 & 11 37 30.07 & $-$62 47 11.8 & 14&530  &    0&464 & $-$0&011 & 0&132  &  0&9589(6)         & EB/EW \\
  v66 & 11 41 42.42 & $-$62 16 32.6 & 14&604  &    0&643 & 0&088 &   0&355  &  2&0593(17)         & EA\\
  v67 & 11 41 57.47 & $-$62 47 56.8 & 14&727  &    0&551 & \multicolumn{2}{c}{---}  &  0&110  &  0&7944(5)      & Ellipsoidal\\
  v68 & 11 43 46.26 & $-$62 42 19.8 & 14&791  &    0&778 & \multicolumn{2}{c}{---}  &  0&110  &  0&46971(16)    & Ellipsoidal\\
  v69 & 11 38 29.28 & $-$62 49 04.3 & 15&602  &    0&598 & \multicolumn{2}{c}{---}  &  0&138  &  0&6902(4)      & Ellipsoidal\\
  v70 & 11 42 56.38 & $-$62 22 34.9 & 15&604  &    0&915 & \multicolumn{2}{c}{---}  &  0&343  &  0&35054(6)     & EW\\
  v71 & 11 42 59.59 & $-$62 19 31.7 & 15&719  &    0&944 & \multicolumn{2}{c}{---}  &  0&408  &  0&9079(7)      & EW\\
  v72 & 11 41 02.00 & $-$62 10 56.7 & 15&742  &    0&836 & \multicolumn{2}{c}{---}  &  0&464  &  \multicolumn{2}{c}{---}      &
EA\\
  v73 & 11 41 08.89 & $-$62 37 51.5 & 15&813  &    0&816 & \multicolumn{2}{c}{---}  &  0&813  &  1&3375(8)      & EA\\
  v74 & 11 43 10.90 & $-$62 23 17.4 & 15&814  &    0&837 & \multicolumn{2}{c}{---}  &  0&192  &  0&39061(6)     & EW\\
  v75 & 11 43 35.82 & $-$62 16 56.8 & 16&047  &    0&729 & \multicolumn{2}{c}{---}  &  0&758  &  0&69449(15)    & EA\\
  v76 & 11 40 35.24 & $-$62 00 33.1 & 16&101  & \multicolumn{2}{c}{---}  & \multicolumn{2}{c}{---}  &  0&589  &  0&36434(12)  &
EW\\
  v77 & 11 41 33.50 & $-$62 16 58.5 & 16&258  &    0&959 & \multicolumn{2}{c}{---}  &  0&143  &  0&8144(9)      & Ellipsoidal/EW\\
  v78 & 11 43 43.72 & $-$62 42 03.3 & 16&614  &    0&751 & \multicolumn{2}{c}{---}  &  0&284  &  0&35965(13)    & EW\\
  v79 & 11 44 00.86 & $-$61 58 16.8 & 16&628  & \multicolumn{2}{c}{---}  & \multicolumn{2}{c}{---}  &  0&527  &  0&36407(12)  &
EA\\
  v80 & 11 42 10.61 & $-$62 27 08.2 & 16&707  &    0&875 & \multicolumn{2}{c}{---}  &  0&388  &  0&342605(11)   & EW\\
  v81 & 11 41 32.87 & $-$62 50 04.5 & 17&039  &    0&655 & \multicolumn{2}{c}{---}  &  0&297  &  0&33349(27)    & EW\\
  v82 & 11 42 09.35 & $-$62 23 58.7 & 17&068  & \multicolumn{2}{c}{---} & \multicolumn{2}{c}{---}  &  0&467  &
\multicolumn{2}{c}{---}     & EA\\
  v83 & 11 43 32.94 & $-$62 44 44.2 & 17&384  & \multicolumn{2}{c}{---} & \multicolumn{2}{c}{---}  &  0&575  &  0&30782(9)    &
EW\\
  v84 & 11 41 05.45 & $-$62 44 17.7 & 17&438  & \multicolumn{2}{c}{---} & \multicolumn{2}{c}{---}  &  0&356  &  0&24675(6)    &
EW\\
  v85 & 11 42 01.93 & $-$62 45 50.3 & 17&485  & \multicolumn{2}{c}{---} & \multicolumn{2}{c}{---}  & 0&754  &  0&32281(8)     &
EW\\
\noalign{\smallskip}\hline\noalign{\smallskip}
       v86 & 11 41 28.09 & $-$62 45 47.8 &  8&975 &     0&322 &  $-$0&005 & 0&049 & \multicolumn{2}{c}{---} & LPV (HD\,101671)\\
       v87 & 11 41 26.52 & $-$62 42 18.4 & 10&882 &     1&791 & \multicolumn{2}{c}{---} & 0&042 & \multicolumn{2}{c}{---} & LPV\\
       v88 & 11 44 20.96 & $-$62 41 23.9 & 11&214 &     0&214 &  $-$0&379 & 0&023 & \multicolumn{2}{c}{---} & LPV\\
$\ast$ v89 & 11 43 04.59 & $-$62 29 19.1 & 13&096 &     1&972 & \multicolumn{2}{c}{---} & 0&102 & \multicolumn{2}{c}{---} & LPV
(W\,400)\\
       v90 & 11 41 43.10 & $-$62 32 49.2 & 13&703 &     2&177 & \multicolumn{2}{c}{---} & 0&106 & \multicolumn{2}{c}{---} & LPV\\
$\ast$ v91 & 11 44 10.57 & $-$62 43 41.1 & 13&734 &     2&089 & \multicolumn{2}{c}{---} & 0&256 & \multicolumn{2}{c}{---} & LPV \\
$\ast$ v92 & 11 44 01.35 & $-$62 14 53.6 & 13&928 &     2&104 & \multicolumn{2}{c}{---} & 0&178 & \multicolumn{2}{c}{---} & LPV \\
$\ast$ v93 & 11 44 13.79 & $-$62 18 15.3 & 14&143 &     2&055 & \multicolumn{2}{c}{---} & 0&132 & \multicolumn{2}{c}{---} & LPV \\
$\ast$ v94 & 11 36 59.97 & $-$62 44 46.9 & 14&207 &     2&476 & \multicolumn{2}{c}{---} & 0&065 & \multicolumn{2}{c}{---} & LPV \\
       v95 & 11 38 52.08 & $-$62 42 40.0 & 14&247 &     1&984 & \multicolumn{2}{c}{---} & 0&130 & \multicolumn{2}{c}{---} & LPV \\
$\ast$ v96 & 11 42 19.00 & $-$62 27 26.6 & 14&262 &     3&293 & \multicolumn{2}{c}{---} & 0&172 & \multicolumn{2}{c}{---} & LPV \\
       v97 & 11 42 21.97 & $-$62 29 50.7 & 14&789 &     0&951 & \multicolumn{2}{c}{---} & 0&107 & \multicolumn{2}{c}{---} & LPV \\
$\ast$ v98 & 11 39 29.31 & $-$62 39 54.4 & 14&795 &     2&504 & \multicolumn{2}{c}{---} & 0&203 & \multicolumn{2}{c}{---} & LPV \\
       v99 & 11 44 11.12 & $-$62 15 52.3 & 15&001 &     0&803 & \multicolumn{2}{c}{---} & 0&142 & \multicolumn{2}{c}{---} & LPV \\
      v100 & 11 41 24.24 & $-$61 59 08.2 & 15&121 &     0&675 & \multicolumn{2}{c}{---} & 0&455 & \multicolumn{2}{c}{---} & LPV \\
      v101 & 11 41 23.37 & $-$62 17 37.7 & 15&192 &     0&747 & \multicolumn{2}{c}{---} & 0&088 & \multicolumn{2}{c}{---} & LPV \\
      v102 & 11 41 25.08 & $-$62 11 13.9 & 16&315 & \multicolumn{2}{c}{---} & \multicolumn{2}{c}{---} & 0&516 &
\multicolumn{2}{c}{---} & LPV \\
      v103 & 11 43 45.74 & $-$62 15 53.7 & 17&527 & \multicolumn{2}{c}{---} & \multicolumn{2}{c}{---} & 0&323 &
\multicolumn{2}{c}{---} & LPV \\
    \noalign{\smallskip}\hline
}

\begin{figure}[!t]
    \includegraphics[width=\hsize]{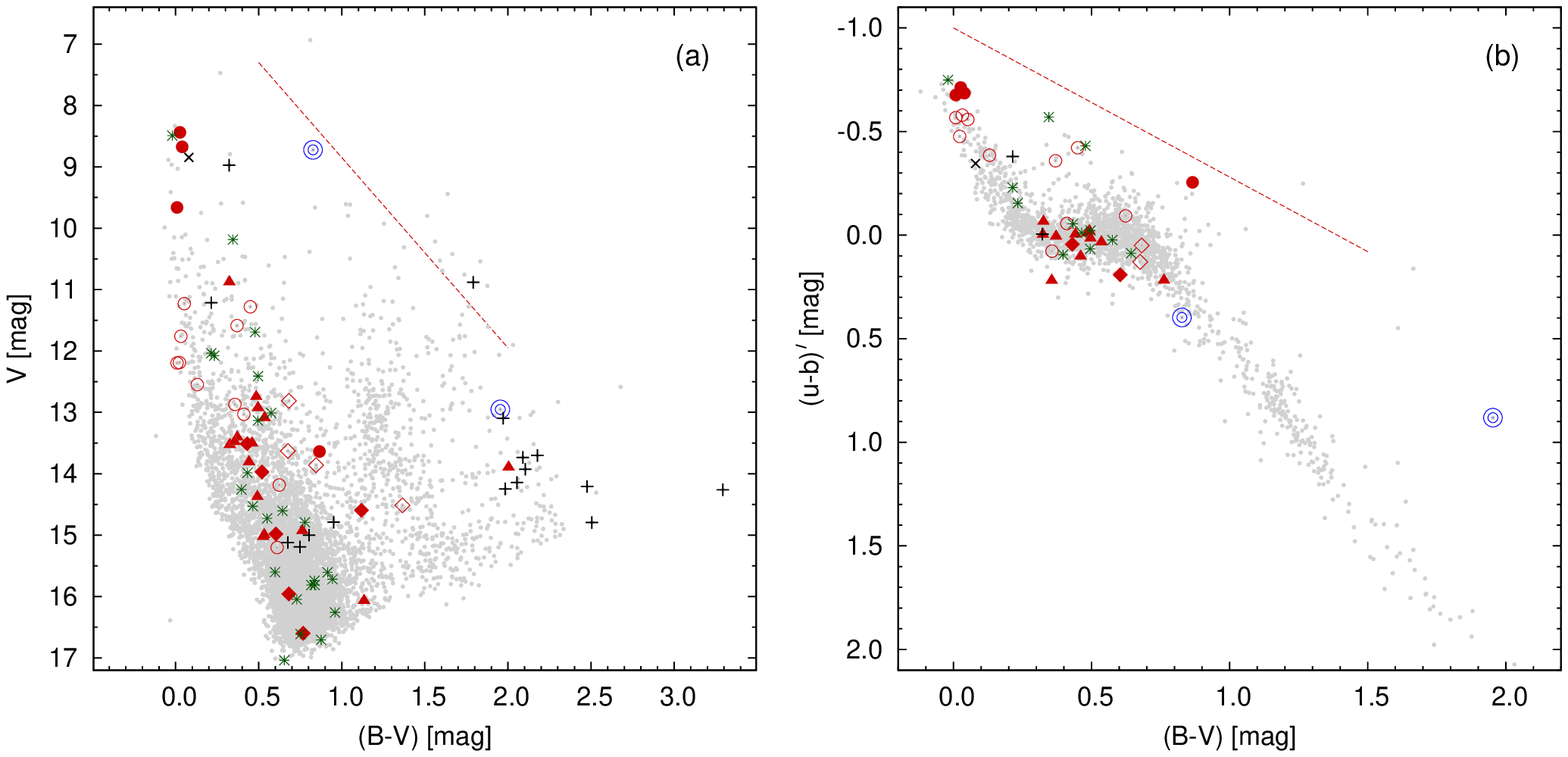}
    \FigCap{{\it Panel (a):} A~collective color-magnitude diagram for stars observed on all five CCD chips of the WFI. The main
sequence of Stock~14 can still be distinguished, although it is now more polluted with field stars than in
Fig.~\ref{fig:cmd_isofit}. The variable stars are denoted with different symbols: $\beta$~Cep stars with filled red circles, SPB
stars (including candidates) with open red circles, $\delta$~Sct stars and candidate monoperiodic stars with red triangles, binary
stars with green asterisks and the LPV stars with plus signs. The filled and open red diamonds stand for monoperiodic stars with
periods shorter and longer than 0.3~d, respectively. The two Cepheids, UZ~Cen and IZ~Cen, are shown with double blue circles.
HD\,102053, an $\alpha^2$~CVn-type variable, is shown with a~cross ($\times$). The reddening line is also shown. {\it Panel (b):}
Two-color $(u-b)^\prime$ vs.~$(B-V)$ diagram for the same stars. Variable stars are shown with the same symbols as in the left
panel.}
    \label{fig:cmd_ccd_var}
\end{figure}

Our survey resulted in finding 103 variable stars in the observed field, 15 of which have already been known as variable in light.
They are listed in Table \ref{tab:var_stars}. The stars were divided into four general variability categories: (i) multiperiodic,
(ii) mono\-pe\-rio\-dic, (iii) eclipsing and ellipsoidal, and (iv) long-period, semi-regular or irregular stars. As far as
sinusoidal light variations are concerned, detection of multiple independent frequencies usually indicates that the variability is
caused by stellar pulsations. When only a~single frequency is detected, one cannot be sure whether this is the case and
alternative explanations should be considered. We assume that the multiperiodic stars presented in Table \ref{tab:var_stars} are
genuine pulsating stars, and attempt to identify their variability type based on the available information, i.e., frequencies,
amplitudes, $UBV$ colors and spectral types. The knowledge of $UBV$ color indices is especially beneficial, as the position of
the star in the two-color diagram can often be used to distinguish the hot $\beta$~Cep and slowly-pulsating B-type (SPB)
variables from cooler $\delta$~Sct and $\gamma$~Dor pulsators. In the previous section, we noted that we do not have reliable
standard $(U-B)$ colors for most of the stars, but even the instrumental color indices are sufficient to this end. Positions of
the discovered variable stars in the color-magnitude and color-color planes are presented in Fig.~\ref{fig:cmd_ccd_var}. The
appearance of those two diagrams will be discussed in detail in Sect.~5. Whenever an~unambiguous identification of the variability
class is impossible, we discuss the possibilities in the text. The parameters of the least squares fits for periodic variables are
not presented here. The table with these parameters and the $UBV$ time series of the discovered variable stars are available from
the \textit{Acta Astronomica Archive}.

In the following sections we shall discuss the basic properties of the stars found to be variable in light. Regarding
identification, we refer to the stars as v1, v2, etc. If a~star appears in the
WEBDA\footnote{\textit{http://www.univie.ac.at/webda}} database, we also provide its WEBDA number, prefixed with~W, in the last
column of Table \ref{tab:var_stars}. The variables can be identified with their equatorial coordinates which are also provided in
Table \ref{tab:var_stars}. The coordinates were calculated in the following way. First, positions of all stars in all $V$-filter
frames were transformed to the positions of a~single arbitrarily chosen reference frame. These positions (in pixels) were averaged
and transformed to equatorial coordinates via astrometric transformation using stars from the USNO\,B1.0 catalog (Monet \etal
2003) as a~reference. This was done separately for each CCD. Typically, about 400 stars in common with the USNO\,B1.0 catalogue
were used to calculate transformation for a~given chip. Standard deviation of these transformations amounted to about 0.25~arcsec.

\subsection{$\beta$~Cephei Stars}
\begin{figure}[!t]
    \includegraphics[width=\hsize]{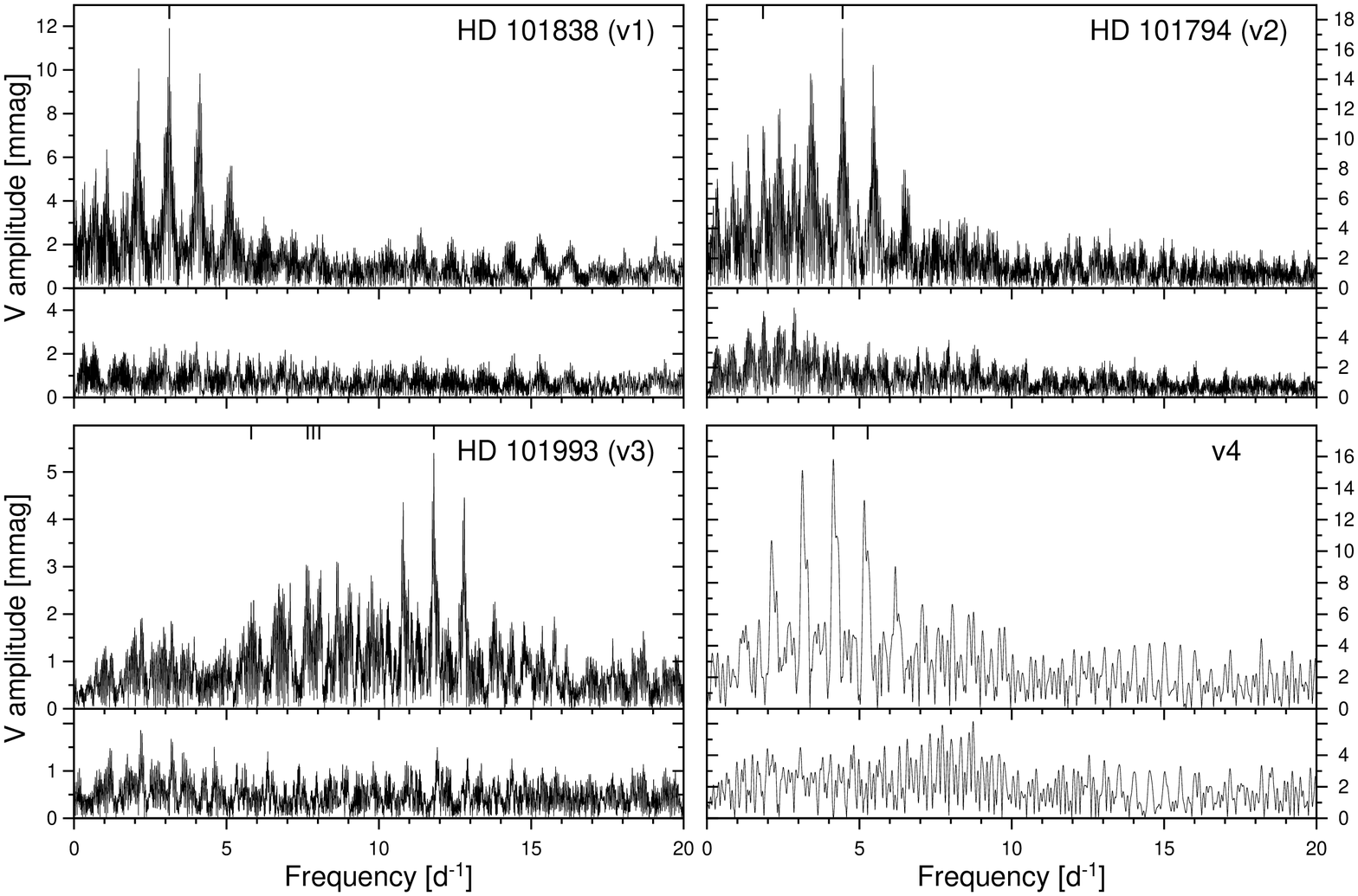}
    \FigCap{Fourier frequency spectra of $V$-filter observations of four $\beta$~Cep stars. For v1 and v2 \textit{top panels} show
the pe\-rio\-do\-gram of the data freed from the contribution of eclipses; for the remaining two stars (v3 and v4) the original
data were used. \textit{Bottom panels} show periodograms of residuals after prewhitening with all significant periodic terms;
their frequencies are marked with tick marks on the top horizontal axis. For clarity, the spectra are shown only up to
20~d$^{-1}$.}
    \label{fig:v1-v4}
\end{figure}

One of the most important results of this work is the confirmation of both the eclipses and low-amplitude variations reported by
Pigulski and Pojma\'{n}ski (2008) in HD\,101838 (v1) and HD\,101794 (v2). The photometry of the two stars was analyzed in the
follo\-wing way. First, the neighboring points in the light curves were averaged to increase the signal-to-noise ratio. Visual
inspection of the resulting averaged time series revealed the presence of eclipses, the times of which were in a~very good
agreement with the ephemeris given by Pigulski and Pojma\'{n}ski (2008). In order to remove the contribution of eclipses, we
assumed the orbital periods derived from the $V$-band ASAS-3 photometry by aforementioned authors, and fitted them to our data
using an appropriate number of harmonics. Then, the residuals were analyzed. In HD\,101838 we found light variations with
a~frequency of $f_1=$ 3.1273(12)~d$^{-1}$. In HD\,101794, we detected two terms with frequencies of $f_1=$ 4.4506(4)~d$^{-1}$ and
$f_2=$ 1.8415(5)~d$^{-1}$ (Fig.~\ref{fig:v1-v4}). In conclusion, we can state that the variability reported by Pigulski and
Pojma\'{n}ski (2008) is confirmed for both targets. However, there is still room for improvement. Removing the contribution of
eclipses by fitting a~sine term with orbital period and its harmonics is simplistic, and the ana\-ly\-sis requires a~more diligent
approach. We have acquired additional photometric and spectroscopic observations of HD\,101838 and HD\,101794, and we expect to do
a detailed light curve modeling in the future, based on all available data. The eclipsing light curves of both stars, this time
freed from the contribution of pulsations, are presented in the first two panels of Fig.~\ref{fig:binaries}. The light curve of
HD\,101838 features a~shallow primary eclipse, and a~hint of what could be a~secondary. In HD\,101794 one can clearly see two
eclipses of similar depth.

We detected short-period variability in another bright star, HD~101993 (v3). From the Fourier analysis, we find five frequencies
present in its power spectrum (Fig.~\ref{fig:v1-v4}). The star shows significant low-frequency noise, likely due to the proximity
of two bright stars in the CCD frame, one of them being the severely overexposed V810~Cen. Detrending was done in a~manner
described above, which helped to some extent. The star was classified as B1\,Ib (Houk and Cowley 1975) and B2\,IV-V (FitzGerald
and Miller 1983). Assuming parameters of Stock~14 in accordance with those derived in Section~3.2, we find the position of v3 in
our color-magnitude diagram (Fig.~\ref{fig:cmd_ccd_var}a) to be consistent with a~B2\,V star. In the two-color diagram
(Fig.~\ref{fig:cmd_ccd_var}b), this star is located in the region occupied by B-type stars, and its reddening is consistent with
the cluster membership. In light of the fact that the star is located within one arcmin from the center of Stock~14, we classify
it as a~$\beta$ Cephei-type variable and a~cluster member.

There is another $\beta$~Cephei-type pulsator in the field, v4. Its position in the color-color diagram
(Fig.~\ref{fig:cmd_ccd_var}b) shows it is an early B-type star. The star seems to be severely reddened: about 0.8~mag more than
Stock~14 in terms of $E(B-V)$. Such amount of reddening along with a~large angular distance from the center of the cluster (36
arcmin) suggest that v4 is a~background object. We have detected two modes in the frequency spectrum of this star
(Fig.~\ref{fig:v1-v4}).

All four $\beta$~Cep stars are plotted with filled red circles in both panels of Fig.~\ref{fig:cmd_ccd_var}. For clarity, we do
not label stars in these plots; their positions can be found using the entries given in Table \ref{tab:var_stars}.

\subsection{SPB and Candidate SPB Stars}
\begin{figure}[!t]
    \includegraphics[width=\hsize]{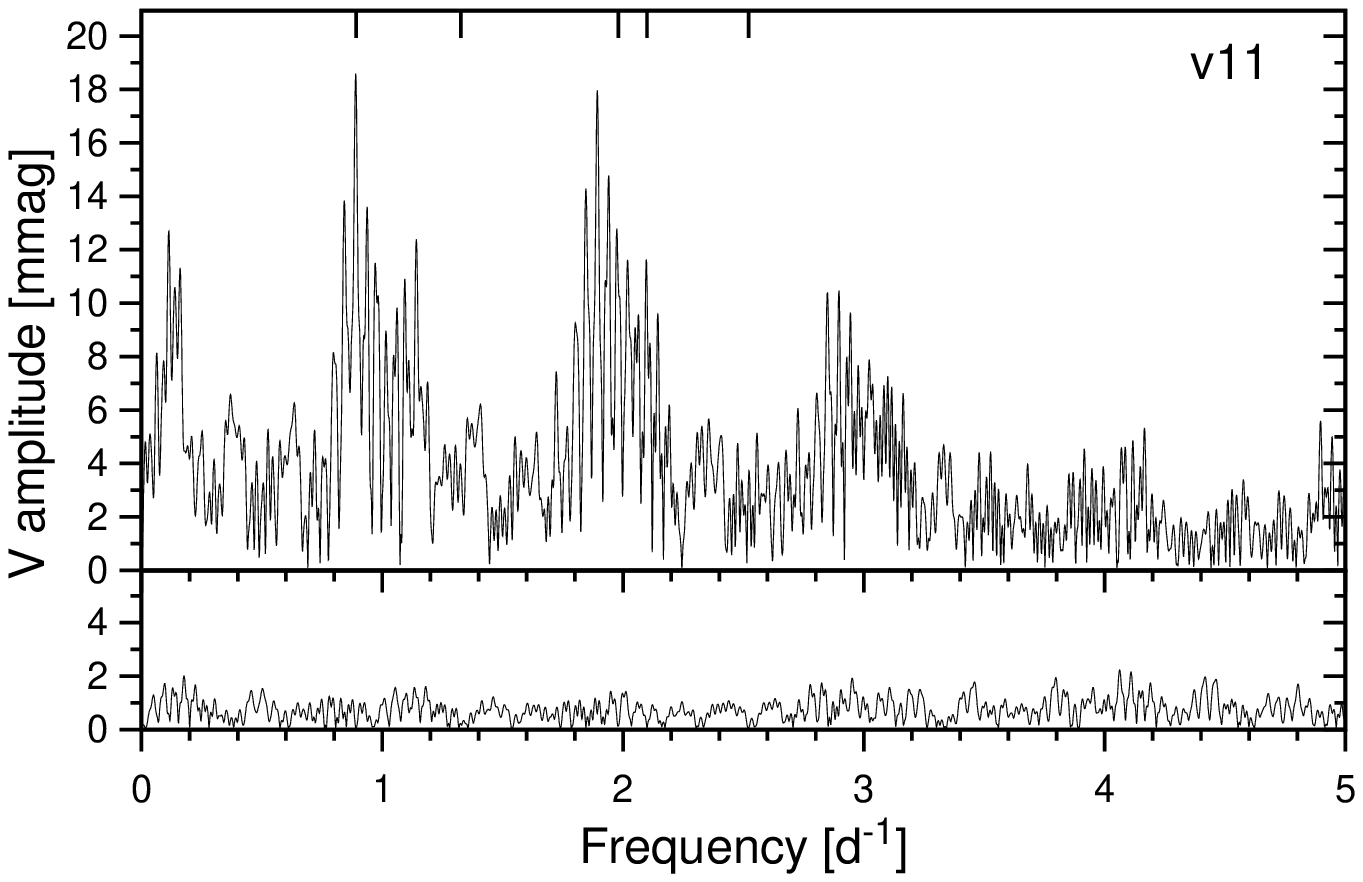}
    \FigCap{Like in Fig.~\ref{fig:v1-v4}, but for the SPB-type star v11. Five frequencies have been found in its Fourier spectrum.
For clarity, the spectrum is shown only up to 5~d$^{-1}$.}
    \label{fig:v11}
\end{figure}

SPB stars have periods in the range between 0.4 and 5~d. Unfortunately, there are other types of periodic variability which
overlap in period with SPB stars. Thus, one can trustworthy classify a~variable as an SPB star only if (i) it is a~B-type star,
and (ii) it is a~multiperiodic variable with periods typical for SPB stars. Unfortunately, there is only one such star in our
sample, v11. Five modes were found in its Fourier spectrum (Fig.~\ref{fig:v11}). The position of v11 in photometric diagrams
(Fig.~\ref{fig:cmd_ccd_var}) indicates that it can be late B spectral type cluster member. This conforms the spectral type B8~V
given by Pickles and Depagne (2010) based on synthetic multi-band photometry. The membership of v11 is also validated by its small
angular distance of less than 4.5~arcmin from the cluster center.
\begin{figure}[!t]
    \includegraphics[width=\hsize]{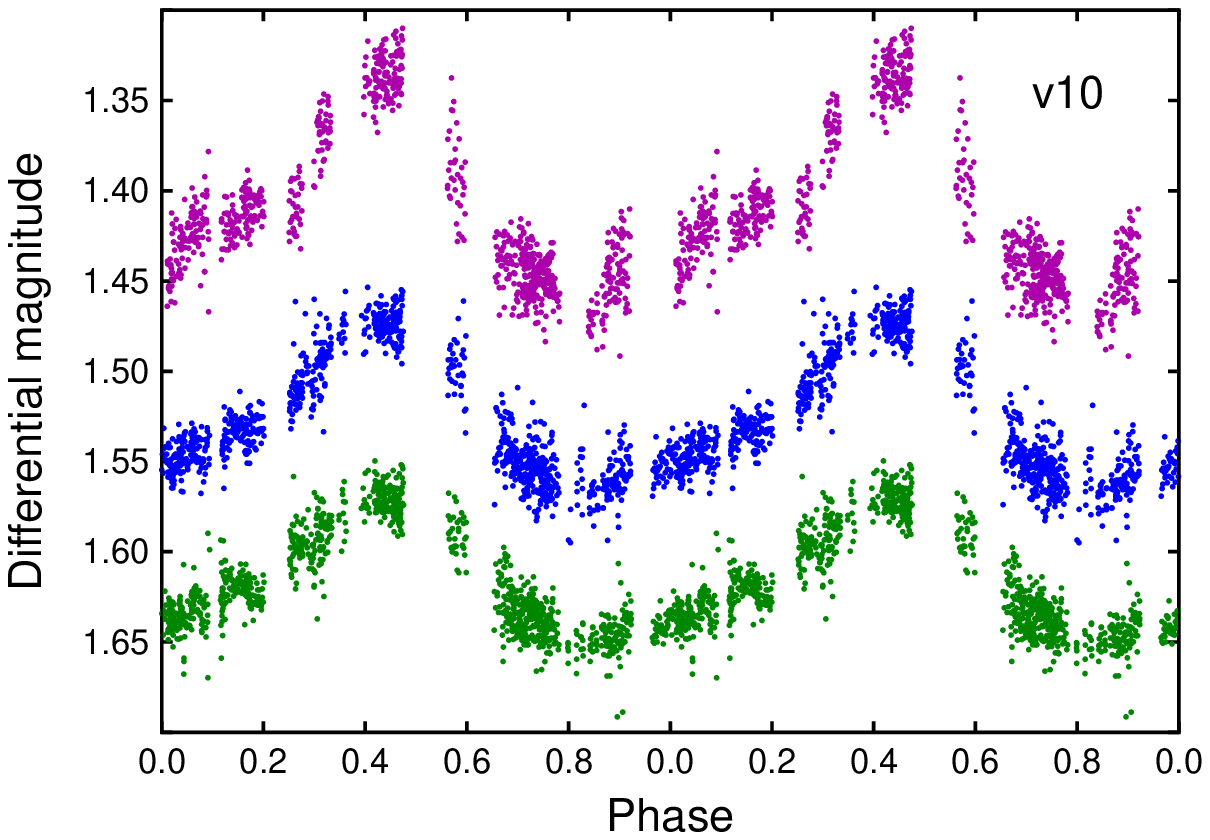}
    \FigCap{$UBV$ light curves of the SPB candidate v10. From top to bottom, $U$, $B$ and $V$-filter observations are plotted,
folded with the period of 3.4108~d.}
    \label{fig:pha_v10}
\end{figure}

There are seven stars (v5 -- v10 and v14) which do not satisfy criterion (ii), i.e., they show only a~single periodicity, but
their periods fall in the range allowable for SPB stars. We therefore consider them to be SPB candidates. Their B spectral type
identity is judged either from direct spectroscopy of a~star or from location in the two-color diagram
(Fig.~\ref{fig:cmd_ccd_var}b). For example, v10 is a~star of spectral type B6\,V (FitzGerald and Miller 1983). The light curves of
this star (Fig.~\ref{fig:pha_v10}) are slightly non-sinusoidal. This does not exclude the possibility that this is an SPB star; an
example of a~similar SPB star is V612~Per = Oo~893 in h~Persei (Krzesi\'{n}ski \etal 1999).

There are three other stars in our sample, which show multiperiodic behavior with periods consistent with SPB classification which
means that they satisfy criterion (ii), but it is not known if they are B-type stars. These are v12, v13 and v15. They could be
SPB stars, but $\gamma$~Dor variability is an alternative. All stars from this group are shown in Fig.~\ref{fig:cmd_ccd_var} with
open red circles.

\subsection{$\delta$~Scuti Stars}
\begin{figure}[!t]
    \includegraphics[width=\hsize]{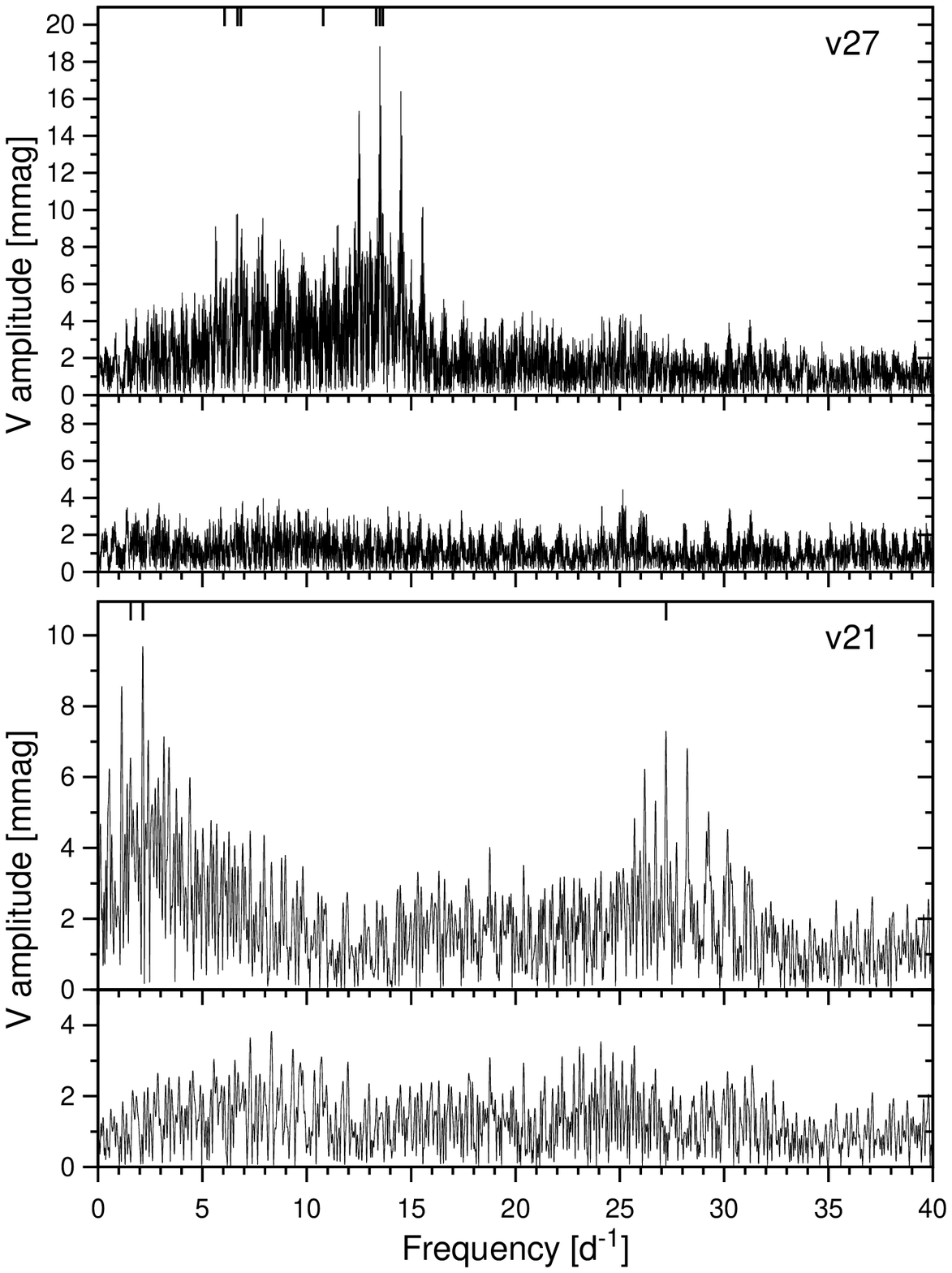}
    \FigCap{The same as in Fig.~\ref{fig:v1-v4}, but for the $\delta$~Scuti-type star v27 (top) and $\delta$~Sct/$\gamma$~Dor
hybrid v21 (bottom). Seven frequencies have been found in the Fourier spectrum of v27; three frequencies, one high and two low in
the spectrum of v21.}
    \label{fig:v27_v21}
\end{figure}

We have discovered nine multiperiodic (v16 -- v18, v20, v22 -- v24, v27 and v28) and four monoperiodic (v19, v26, v29 and v30)
stars with frequencies typical for $\delta$~Scuti-type pulsating stars. The Fourier spectra of multiperiodic stars typically show
two or three modes, but three stars stand out in this respect: v16, v24 and v27 feature four, five and seven modes, respectively.
As an example, the amplitude spectrum of v27 is presented in Fig.~\ref{fig:v27_v21}. The brightest star, v16 (HD~308900), features
four frequencies between 3.89 and 4.97~d$^{-1}$ in its power spectrum. These are low, but still within the theoretical allowed
frequency range for $\delta$~Scuti pulsators. The frequencies of the four single-mode $\delta$~Scuti stars are high enough (higher
than 10~d$^{-1}$) to be sure that there is little probability that their variability is of different origin. Therefore, we
classify these stars as $\delta$~Scuti-type pulsators. The $\delta$~Scuti-type stars with both colors measured occupy positions
consistent with spectral types A to F in the color-color plane. On the other hand, from the color-magnitude diagram one can infer
that they are redder than cluster stars. In consequence, they are likely field objects.

The star v21 exhibits light variations with one high and two low frequencies (Fig.~\ref{fig:v27_v21}). It has a~close neighbor of
similar brightness in the CCD frame. We calculated Fourier transform of the time-series data of the neighbor, to check if it
contributes to the low-frequency noise. This turned out not to be the case. In the color-color diagram this star is situated
slightly below the clustering of points corresponding to A-type stars. This is an instrumental effect: this star has been observed
with the WFI CCD \#4, which is the worst in terms of quantum efficiency in the ultraviolet. In the color-magnitude plane the star
is situated near the discovered $\delta$~Sct pulsators. We conclude that the low frequencies seem to be real, and that the
observed light variations in this star can be interpreted as arising from g-modes typical for $\gamma$~Doradus-type stars and
a~p-mode typical for $\delta$~Scuti-type variables. This makes v21 a~hybrid pulsator. The star seems to be a~field object.

There is a~single multiperiodic star, v25, with three relatively low-amplitude high frequencies detected, 4.0, 7.9, and
7.6~d$^{-1}$. Surprisingly, it is very red, $(B-V) \approx$~2.0. If one assumes a~reasonable behavior of reddening with distance,
its position in the color-magnitude diagram cannot be agreed neither with $\beta$~Cep nor $\delta$~Sct type of variability. The
star has rather poor photometry, and it is possible that its variability is spurious or of instrumental origin. Another
possibility is that the star is severely contaminated by an unresolved very red object. The star v25, the hybrid v21 and all 13
$\delta$~Scuti stars are plotted in Fig.~\ref{fig:cmd_ccd_var} with filled red triangles.

\subsection{Cepheids} \label{sec:cepheids}
\begin{figure}[!t]
    \includegraphics[width=\hsize]{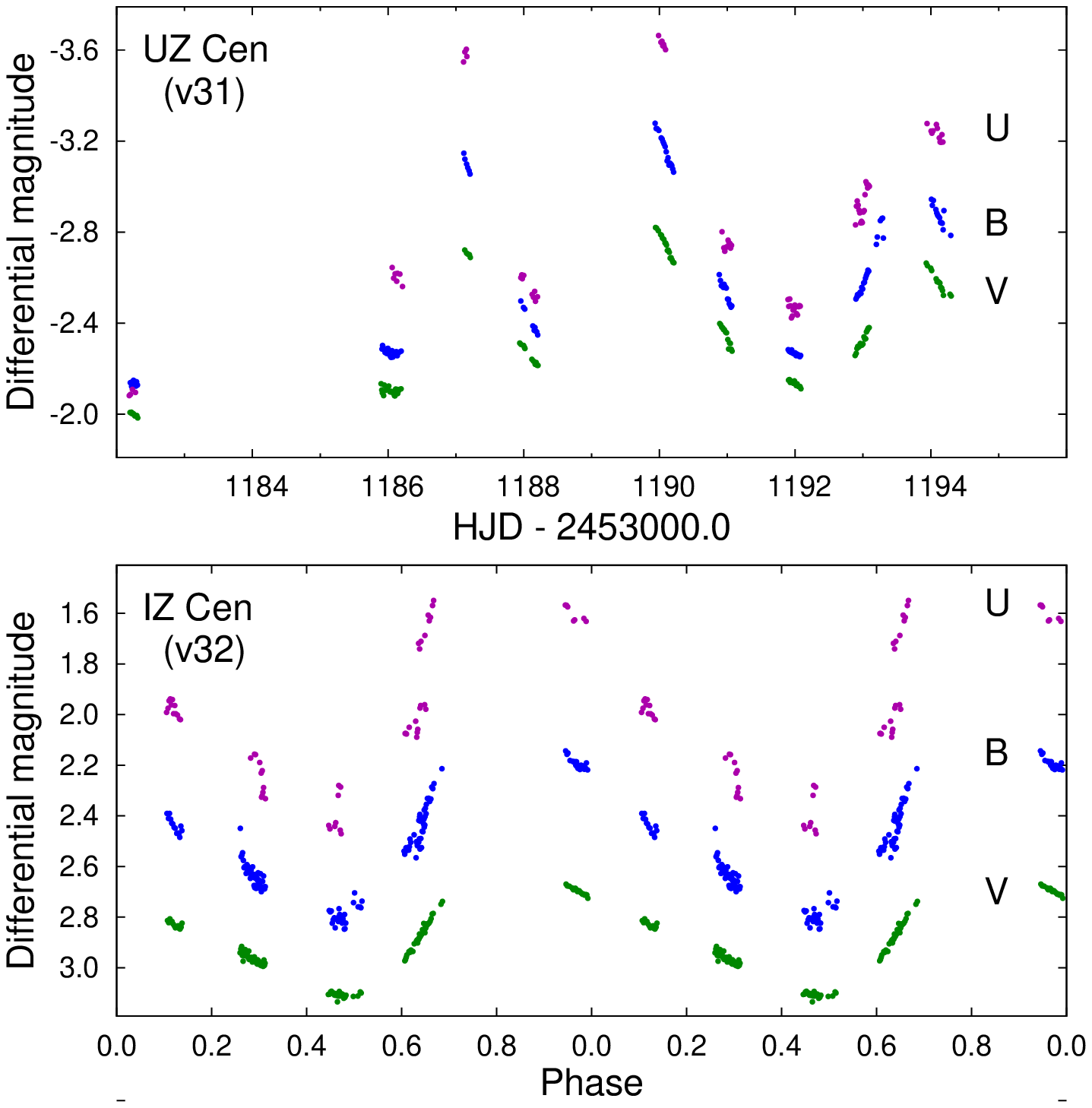}
    \FigCap{Light curves of two known Cepheids, UZ~Cen (top) and IZ~Cen (bottom). The data for IZ Cen were folded with the period
of 5.89237~d.}
    \label{fig:cepheids}
\end{figure}

Two known classical Cepheids, the double-mode UZ~Cen (v31, Pel 1976, Stobie 1976), and IZ~Cen (v32, Walraven \etal 1958), are
located in the field we observed. Their variability is clearly visible in the light curves, but we were unable to recover the
pulsation periods found in the literature due to the incomplete phase coverage. For this reason, we used the ASAS-3 data to derive
accurate periods of both stars. We obtained 3.334311(3)~d and 2.355204(7)~d for the fundamental and first overtone radial mode of
UZ~Cen, and 5.89237(21)~d for IZ~Cen. The light curves of both stars in three passbands are presented in Fig.~\ref{fig:cepheids}.
Since the WFI time series of UZ~Cen could not be decomposed, its light curve is plotted as a~function of time, whereas IZ~Cen was
folded with its ASAS-3 period. The two Cepheids are plotted in Fig.~\ref{fig:cmd_ccd_var} with double open blue circles.

\subsection{Monoperiodic Stars}
The next twenty two variable stars form a~group of periodically variable stars which, with one exception, cannot be unambiguously
classified. The exception is bright star v33 (HD~102053) which shows variability with a~period of about 11.4~d and peak-to-peak
amplitude of about 0.09~mag in $V$ filter. This is a~known chemically peculiar star classified as Ap~Si by Bidelmann and MacConnel
(1973) and Houk and Cowley (1975) and B8\,Vp (Si) by FitzGerald and Miller (1983). Taking into account spectral peculiarities, the
observed light variations can be interpreted in terms of rotational variability caused by spots in the stellar photosphere. We
therefore classify HD\,102053 as an $\alpha^2$~CVn-type variable star. In the color-magnitude diagram
(Fig.~\ref{fig:cmd_ccd_var}a) the star is situated near the top of the main sequence, in the region occupied by early B-type
stars; it is plotted with a~cross ($\times$). This suggests that v33 is a~foreground object as already concluded by FitzGerald and
Miller (1983).

The remaining monoperiodic variable stars  presented in Table~\ref{tab:var_stars}, v34 -- v54, have periods and amplitudes in
a~range from 0.1 to 3.8~d and amplitudes up to 0.5~mag. Most of them are faint, and we cannot exclude the possibility that some of
them are pulsating. For example, the stars with short periods and small amplitudes (e.g., v35 and v38) might be pulsating stars of
$\delta$~Sct-type. Those with relatively short periods and large amplitudes could be in turn low-inclination W~UMa-type binary
systems. Rotation and surface inhomogeneities or proximity effects in a~binary system may also cause variability in stars included
in this group. A good candidate for rotationally-induced variability is v39 which has non-sinusoidal light curve. Except for v33,
the monoperiodic stars are plotted in Fig.~\ref{fig:cmd_ccd_var} with filled and open red diamonds, respectively for stars with
periods shorter and longer than 0.3~d.

\subsection{Eclipsing and Ellipsoidal Systems}
\begin{figure}[htb]
    \includegraphics[width=\hsize]{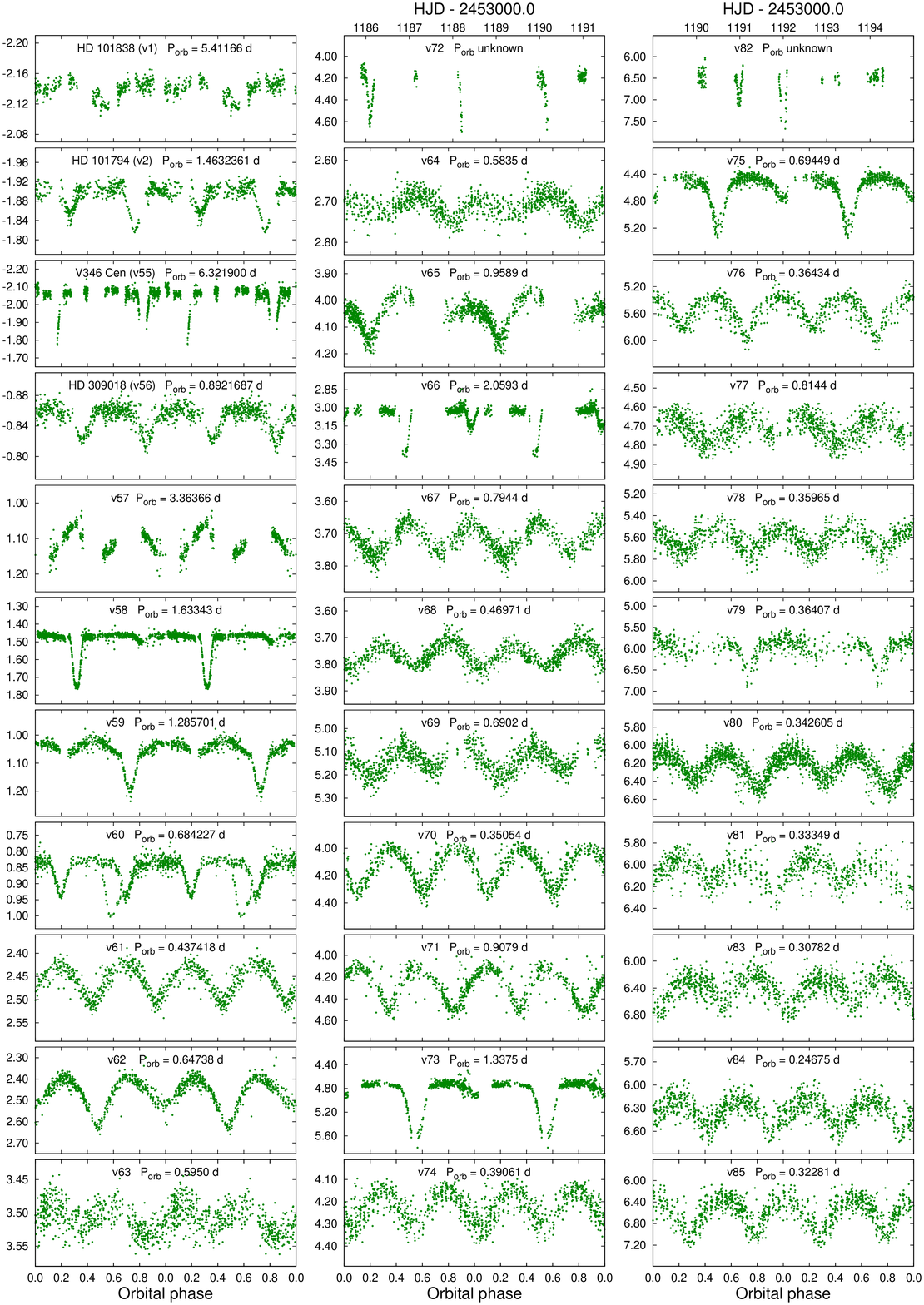}
    \FigCap{Differential $V$-filter light curves of the 33 binary systems. In all but the two upper panels in central and right
column the abscissae show the doubled orbital phase.}
    \label{fig:binaries}
\end{figure}

Apart from HD\,101794 and HD\,101838 described in Section 4.1, our survey resulted in detecting 31 eclipsing and ellipsoidal stars
including 12 EA-type systems, one EB-type system, 11 EW-type systems and 7 ellipsoidal variables. The light curves of all 33
eclipsing and ellipsoidal systems are shown in Fig.~\ref{fig:binaries}. Four eclipsing binaries have been known as variable prior
to this study: v1, v2, v55 = V346~Cen (O'Leary and O'Connell 1936, Hern\'{a}ndez and Sahade 1978, Gimen\'{e}z \etal 1986) and v56
= HD~309018 (Pojma\'{n}ski 1998). The magnitudes and colors in Table \ref{tab:var_stars} correspond to the phase outside of
eclipses. In this section we comment on the most interesting findings and devote a~separate subsection only to V346~Cen.

In our photometry, v60 appeared to be variable with a~period of 0.6855 d. After folding the time series data with that period, we
obtained a~light curve showing two eclipses of a similar depth. Apart from that, an additional deeper minimum is seen around phase
of 0.6 (Fig.~\ref{fig:binaries}). This minimum was observed only once, on HJD\,2454187.17. At $V=$ 12.41~mag the star is quite
faint for the ASAS-3 telescopes, nevertheless the analysis of available ASAS-3 data resulted in finding a~more accurate value of
$P_{\rm{orb}}=$ 0.684227(4)~d. The deeper minimum is not visible in the ASAS-3 light curve. Its origin is not clear. The most
plausible explanation is that our photometry is contaminated by an unresolved eclipsing binary system.

Another interesting system is v65. While tentatively classified as an EW-type binary star, its orbital period of 0.9589(6) d is
quite long for a~W~UMa-type system. Folded with this period, the light curve exhibits the~primary eclipse, and a~gap in the data
where the secondary should be. What is interesting, however, is the difference in out-of-eclipse brightness on both sides of the
gap. This could indicate that a~significant O'Connell effect is present in this binary system. More observations would be
necessary to achieve complete phase coverage, which would make light curve modeling possible.

Judging from Fig.~\ref{fig:cmd_ccd_var}, v56 and v57 are of early spectral type. The star v57 has a~gapped light curve in our
data, but the analysis of the ASAS-3 data allowed for unambiguous determination of its period (3.36366 d). It seems to be an
ellipsoidal variable. The EA-type eclipsing variable v57 (HD\,309018, CPD $-$62$^{\circ}$2261) is classified as O9\,V
(Humphreys 1973). Both v56 and v57 are more reddened than Stock 14; we discuss the possibility that they belong to Cru~OB1 in
Sect.~5.

For v72 and v82 we could not derive the orbital periods because the data probably cover only their secondary minima and possibly
fragments of the primary ones. Unfortunately, attempts to find a~period assuming circular orbit failed, so that these are likely
systems with eccentric orbits and relatively long orbital periods. More observations are necessary to improve the phase coverage
and to calculate the orbital periods of those two systems.

\subsection{V346~Cen = HD\,101837}
V346~Cen (v55) was discovered as an eclipsing variable with unevenly spaced minima of similar depth by O'Connell (O'Leary and
O'Connell 1936). This prom\-pted Dugan and Wright (1939) to study the star using Harvard College Observatory regular patrol plates
dating back to 1890. They were the first to show the rotation of the line of apsides in the binary, a~result confirmed later by
O'Connell (1939) and Gim\'{e}nez \etal (1986). The period of apsidal motion derived by the latter authors amounted to 321 $\pm$
16~yr. They only used the times of minima obtained by themselves and O'Connell (1939).

\begin{figure}[htb]
    \includegraphics[width=\hsize]{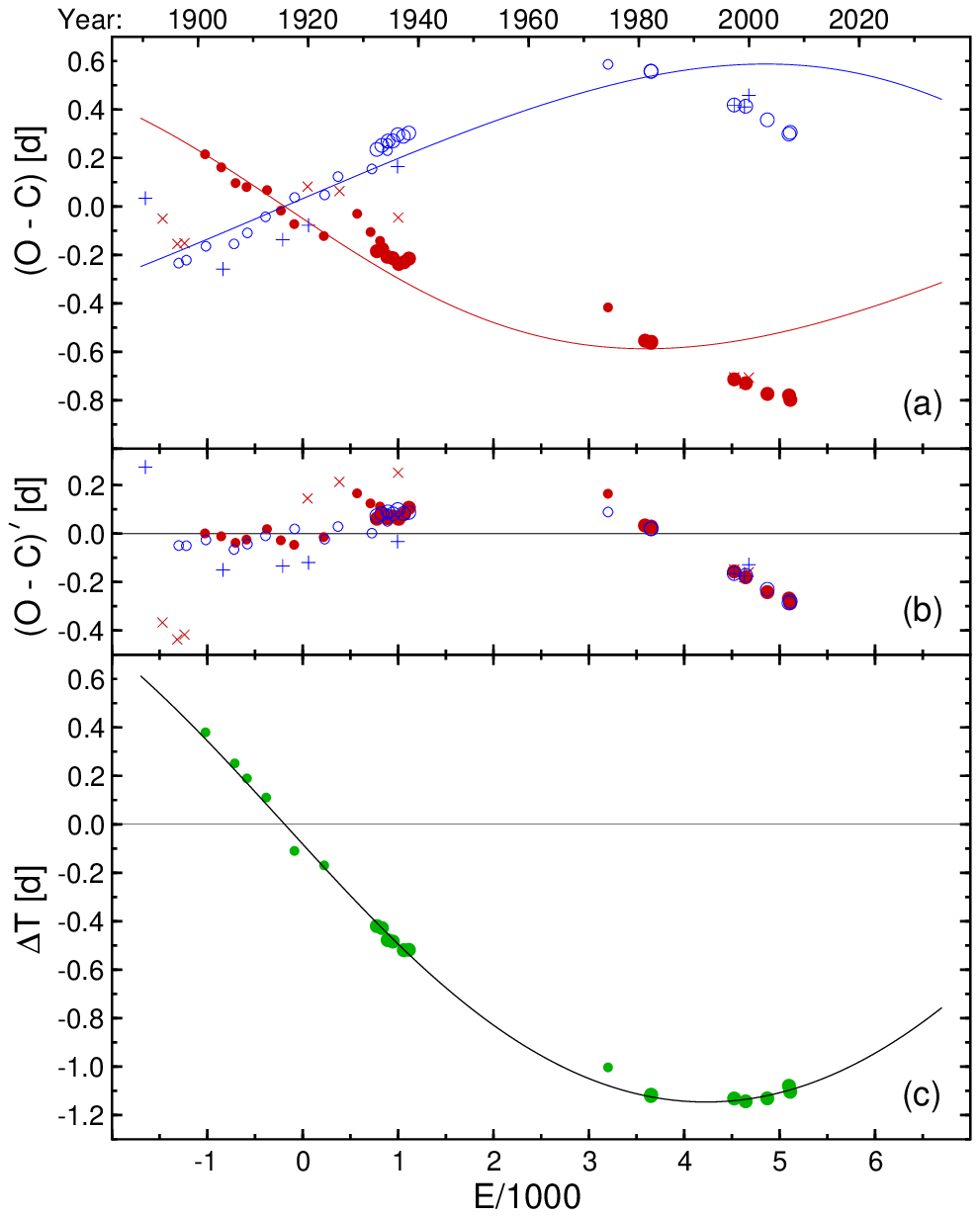}
    \FigCap{{\it Top:} $O-C$ diagram for the times of primary (filled red circles) and secondary (open blue circles) minima of
V346~Cen. Larger symbols are used for data which were given higher weights in our fits. Times of minima listed in the literature
but not used in our fits are shown with crosses (primary) and plus signs (secondary). {\it Middle:} The same as in panel (a), but
after subtracting the changes due to apsidal motion.  {\it Bottom:} Time spans between secondary and primary minima, $\Delta T$,
as defined in the text. Larger symbols stand for data with higher weights. Continuous line is the fit.}
    \label{fig:o-c}
\end{figure}

With the new data, including the ASAS observations and our own photometry, we have revised period changes in the V346~Cen system.
First of all, we derived times of minimum light for four sets of data that became available since the analysis of Gim\'{e}nez
\etal (1986): the ASAS-2 $I$-filter (Pojma\'{n}ski 1998, Pojma\'{n}ski 2000) and ASAS-3 $V$-filter data, our own $UBV$ photometry
and 1974 Walraven $VBLUW$ photometry published recently by van Houten \etal (2009). The ASAS-2 and ASAS-3 data were divided into
two subsets. Due to the scarceness of the data, for each set we folded the magnitudes with the appropriate period and then fitted
eclipses by two Gaussian functions. The resulting phases of minimum light were converted to times of minimum adjacent to the mean
epoch of a~given subset. The calculated times of minimum light are listed in Table \ref{tab:V346}. We note that due to the similar
depth of eclipses, there is confusion in the literature regarding their naming. If following Gim\'{e}nez \etal (1986) we name the
slightly deeper eclipse the primary one, then the opposite naming is adopted by Dugan and Wright (1939), Kreiner \etal (2001) and
Og\l{}oza and Zakrzewski (2004). The new $O-C$ diagram for V346~Cen is presented in Fig.~\ref{fig:o-c}a. The $O-C$ diagram was
calculated with the following ephemeris:
\[ \mbox{Min I} = \mbox{HJD 2421960.513} + \mbox{6.322046}^{\rm d} \times E \]
and
\[ \mbox{Min II} = \mbox{HJD 2421963.674} + \mbox{6.322046}^{\rm d} \times E, \]
where $E$ is the number of orbital periods elapsed from the initial epoch.  Our ephemeris for the secondary minimum is exactly the
same as used by Kreiner \etal (2001) for the eclipse they named ``primary''.

\MakeTable{l r@{.}l r@{.}l}{12.5cm}{\label{tab:V346}New times of minimum light for V346~Cen.}
{
\hline
          & \multicolumn{4}{c}{Time of minimum light} \\
Source of data & \multicolumn{4}{c}{(HJD\,2\,400\,000.0+)} \\
      & \multicolumn{2}{c}{Primary} & \multicolumn{2}{c}{Secondary} \\
\hline
van Houten \etal (2009) & 42196&965(06)  & 42201&130(10) \\
ASAS-2, part 1   & 50560&7355(53) & 50558&7066(57)\\
ASAS-2, part 2   & 51325&6870(23) & 51323&6692(23)\\
ASAS-3, part 1   & 52760&7474(32) & 52758&7175(42)\\
this work, $V$-filter  & 54195&8452(66) & 54193&7633(20)\\
ASAS-3, part 2   & 54278&0140(20) & 54275&9568(43)\\
\hline
}

It is clear from Fig.~\ref{fig:o-c}a that a~sole apsidal motion cannot explain the changes seen in the $O-C$ diagram and that
another effect must be involved. The most plausible candidate is the light-time effect in a~wide system, which affects the times
of both minima in the same way. In order to separate the effects we decided to use the time span between the primary and
secondary minima, as this value is not affected by the light-time effect. The time spans, $\Delta T$, in the sense $\Delta T =
(\mbox{Min I}) - (\mbox{Min II}) + P_{\rm a}/\mbox{2}$ were calculated for 20 pairs of minima that were separated by less than 20
orbital periods. $P_{\rm a}$ stands for the anomalistic period. The resulting values of $\Delta T$ are plotted in
Fig.~\ref{fig:o-c}c. Following the formalism presented by Gim\'{e}nez and Garcia-Pelayo (1983) and Gim\'{e}nez and Bastero (1995),
the time span defined as above equals to
$$\Delta T = -A_1\frac{eP_{\rm a}}{\pi}\cos\omega + A_3\frac{e^3P_{\rm a}}{4\pi}\cos3\omega - A_5\frac{e^5P_{\rm a}}{16\pi}\cos
5\omega,$$
where $e$, $P_{\rm a}$, and $\omega$ denote eccentricity, anomalistic period, and longitude of periastron, respectively. $A_1$,
$A_3$ and $A_5$ are coefficients defined in the cited papers. They depend on the inclination, $i$, and eccentricity. In the
presence of apsidal motion, the only parameter that varies with time is $\omega$; the variability is assumed to be linear, $\omega
(E) = \omega_0 + \dot\omega\times E$, where $\omega_0 = \omega(0)$ and $\dot\omega$ is the angular velocity of the line of
apsides. Fixing the period and assuming $i = \mbox{83.9}^{\circ}$ (Gim\'{e}nez \etal 1986), the only parameters that can be
derived are $\omega_0$, $\dot\omega$ and $e$. They define respectively the phase, period and amplitude of the changes of $\Delta
T$. From the fit made by means of non-linear least squares we obtained $\omega_0 = \mbox{273.87}^{\circ}$ $\pm$
$\mbox{0.43}^{\circ}$, $\dot\omega = \mbox{0.02039}$ $\pm$ $\mbox{0.00027}~^{\circ}$/sidereal period, and $e = \mbox{0.2873}$
$\pm$ $\mbox{0.0010}$. The uncertainties of parameters were derived by means of the bootstrap method.

The eccentricity we obtained is in a~very good agreement with the value of 0.288 $\pm$ 0.003 provided by Gim\'{e}nez \etal (1986).
From $\omega_0$ and $\dot\omega$ we can calculate the epoch when the minima were separated by half the period ($\omega =
\mbox{270}^{\circ}$). It is 1915.7 $\pm$ 0.2, with a~fair agreement with previous determinations. Finally, the period of apsidal
motion $U = \mbox{360}^{\circ} P_{\rm s}/\dot\omega =$ 306 $\pm$ 4 yr, where $P_{\rm s}$ is the sidereal period. The $U$ is within
1$\sigma$ of the value obtained by Gim\'{e}nez \etal (1986). However, we believe that our value is much better constrained,
because the data we used span twice as long as those used by Gim\'{e}nez \etal (1986). In addition, to derive $U$ we used values
of $\Delta T$ which are free from the suspected light-time effect affecting the $O-C$ diagram.

Having obtained the three parameters, we have subtracted the changes in the $O-C$ diagram from the contribution of the apsidal
motion. The resulting diagram is shown in Fig.~\ref{fig:o-c}b. It can be seen that the residual changes, $(O-C)^\prime$, can be
indeed due to the light-time effect. Unfortunately, the data do not cover the whole orbital period of the wide system and we can
only conclude that this period is longer than 100 yr.

\subsection{Long-Period Variables}
Our study revealed the presence of 18 stars (v86 -- v103) which are clearly variable in light, but it was not possible to
determine whether the changes are periodic or not. We classify these stars as long period variables (LPVs). Seven stars of this
sample have been previously known to be variable. They were classified as ``miscellaneous'' variables in the ASAS-3 catalogue
(Pojma\'{n}ski 2000). The magnitudes and color indices for stars in this group are given for the epoch of HJD\,2454193.1. They
were calculated from the average $UBV$ magnitudes acquired on that night. The variability range calculation was also based on the
overnight averages, and represents the difference between the faintest and the brightest observed mean values. Two stars in the
sample, v86 and v88, are relatively blue. Judging from its position in the two-color diagram, v88 is a~B-type star. The other
one, v86 (HD~101671) is classified as A0\,Vm: (Houk and Cowley 1975).

% ===============================================================================================================================
% ===============================================================================================================================
\section{Discussion and Conclusions}
We have confirmed the low-amplitude light variations detected by Pigulski and Pojma\'{n}ski (2008) in HD\,101838 (v1) and
HD\,101794 (v2). These variations were interpreted by the authors as arising from $\beta$~Cephei-type pulsations. The frequency of
about 3.13~d$^{-1}$ found in v1 is close to the low-frequency limit for p-modes in these stars (Pamyatnykh 1999). However,
a~g-mode or a~mixed mode is a~possibility. HD\,101794 is an even more interesting case: two low-amplitude periodic terms are seen,
one with frequency in the range expected of $\beta$~Cephei stars, and another one which could arise from a~g-mode (SPB-type
pulsations). If both originated in the same component, it would make this star a~hybrid pulsator, a~very valuable candidate for
seismic modeling. There is another possibility, however. The star is classified as a~Be star (Henize 1976, Garrison \etal 1977).
In light of this, the variations with the lower frequency might not be caused by stellar pulsations, but instead be of
$\lambda$~Eri-type seen in the component with Balmer lines in emission. The question of the origin of the observed variations
could in principle be answered by the analysis of in-eclipse photometry. However, this would require photometric quality superior
to that of the WFI data.

Fortunately, we have discovered another $\beta$~Cephei-type member of Stock~14, HD\,101993 (v3). This makes Stock~14 a~promising
object for the application of ensemble asteroseismology. The fact that stellar parameters for two of the three $\beta$~Cephei-type
pulsators could be derived from the combination of the eclipsing light curve and radial-velocity measurements, is an additional,
very strong constraint. The seismic modeling of this kind, however, is beyond the scope of this paper. It is postponed to
a~separate paper where additional multi-band photometry and spectroscopy will be used.

The yield of the present survey is a~list of 103 variable stars, 88 of which have not been known as variable prior to this study.
The majority of the discovered variable stars are field stars, not physically associated with Stock~14. This is supported by the
fact that in the color-magnitude plane most of them form a~strip coincident with the sequence of field stars. Judging from the
photometric diagrams, only ten variable stars can be regarded as members of Stock~14. In addition to the three
$\beta$~Cephei-type stars (v1, v2, v3), these are: the SPB star v11, four SPB candidates (v5, v8, v9, and v10), and the eclipsing
binary V346~Cen (v55). The B-type LPV v88 is the tenth possible member provided it is slightly more reddened than the main bulk of
cluster members. Of course, a~possibility that some faint variable stars, especially the eclipsing and ellipsoidal ones, also
belong to Stock~14, cannot be excluded.

In the color-color diagram (Fig.~\ref{fig:cmd_ccd_var}b) we note the existence of a~group of about ten OB-type stars reddened
approximately 0.3~mag more than the cluster. The group includes four variable stars: two SPB candidates v6 and v7, the eclipsing
O9\,V-type star HD\,309018 (v56) and the likely ellipsoidal variable v57. There is a~possibility that these stars belong to the
Cru~OB1 association that surrounds Stock~14 in the sky. The association was defined by Klare (1967) as K12, and later studied
e.g.~by Lyng\aa{} (1970), Humphreys (1976, 1978) and Kaltcheva and Georgiev (1994). There are, however, other investigations
indicating the presence of many more young stellar complexes in this direction (Mel'nik and Efremov 1995, Tovmassian \etal 1996a,
1996b, 1998). This is not unexpected, as the line-of-sight in the observed direction crosses the Sagittarius-Carina spiral arm up
to the distance of 6~kpc. In any case, Cru~OB1 contains stars of as early spectral type as O6\,V, so that it is much younger than
Stock~14. It is also slightly more distant than Stock~14 which is located at the outer edge of the spiral arm. The members of
Cru~OB1 are therefore more reddened than Stock~14 with $E(B-V)$ reaching 0.5~mag, which agrees with the reddening of the group of
stars we mentioned above. The two brightest stars seen in Fig.~\ref{fig:cmd_ccd_var}a are the resolved components of the O9.5~Ib
binary HD~101545 and are also members of Cru~OB1 (Humphreys 1978, Kaltcheva and Georgiev 1994).

The mean reddening and distance of the Stock~14 cluster presented in this work differ from the values given by Moffat and Vogt
(1975), Turner (1982), Peterson and FitzGerald (1988) and Kharchenko \etal (2005). Our results suggest the cluster is slightly
less reddened: 0.21~mag as opposed to about 0.25~mag in terms of $E(B-V)$. With the distance of 2.4~kpc, it also seems to be
situated further away from the Sun. The differences are likely caused by using a~different intrinsic color relations for main
sequence stars and more recent isochrones including more accurate treatment of stellar physics. The fact that our photometric
survey is much deeper than the previous ones suggests our results can be regarded as more precise.

% ===============================================================================================================================
% ===============================================================================================================================
\Acknow{The authors wish to thank the Director of the Research School of Astronomy and Astrophysics of the Australian National
University for time on the 40-inch telescope at Siding Spring Observatory. This research has made use of the WEBDA database,
operated at the Institute for Astronomy of the University of Vienna, the Aladin sky atlas and VizieR catalogue access tool
operated at the CDS, Strasbourg, France, and the SAOImage DS9, developed by the Smithsonian Astrophysical Observatory. This work
was supported by the funds granted by the National Science Centre, on the basis of decision DEC-2011/01/N/ST9/00400. Calculations
have been carried out in Wroc{\l}aw Centre for Networking and Supercomputing (\textit{http://www.wcss.wroc.pl), grant No.~217.}

% ===============================================================================================================================
% ===============================================================================================================================

\end{document}